\documentclass{article}

\PassOptionsToPackage{numbers}{natbib}

\usepackage[preprint]{neurips_2026}

\usepackage[utf8]{inputenc} 
\usepackage[T1]{fontenc}    
\usepackage{hyperref}       
\usepackage{url}            
\usepackage{booktabs}       
\usepackage{amsfonts}       
\usepackage{nicefrac}       
\usepackage{microtype}      
\usepackage{xcolor}         
\usepackage{graphicx}
\usepackage{amsmath}
\usepackage{mathtools}
\usepackage{cleveref}
\usepackage{multirow}
\usepackage{wrapfig}

\usepackage{algorithm}
\usepackage{algorithmic}
\usepackage{comment}

\title{A meta-algorithm for ab initio reconstruction of complex mixtures in cryo-EM}

\author{%
  Alkin Kaz \\
  Princeton University\\
  \texttt{akaz@princeton.edu} \\
  \And
  Arda Kaz \\
  University of California, Berkeley \\
  \texttt{ardakaz@berkeley.edu} \\
  \And
  Ellen D.~Zhong \\
  Princeton University \\
  \texttt{zhonge@princeton.edu} \\
}

\begin{document}

\maketitle

\begin{abstract}
    We describe a systematic approach for spawning and aggregating multi-class cryo-EM reconstruction jobs. This approach formalizes standard ad hoc strategies of iterative classification and filtering typically used by practitioners to sort impure, heterogeneous samples. To our knowledge, this is the first method that can successfully perform \textit{ab initio} reconstruction on datasets containing dozens of distinct species. We obtain 97\% accuracy on \textit{ab initio} reconstruction of a 45-class subset of Tomotwin-100, 75\% accuracy on the full Tomotwin-100 dataset, and demonstrate recovery of ribosomal assembly states from an unfiltered experimental cryo-EM dataset. Our approach's capability scales with compute and lays the foundation for automated cryo-EM workflows in modern experimental settings.
\end{abstract}

\section{Introduction}

Cryo-electron microscopy (cryo-EM) is a central tool for determining the 3D structures of biological molecules at near-atomic resolution. In a typical single particle cryo-EM experiment, an aqueous solution of molecules is vitrified, imaged as randomly-oriented 2D projections by a transmission electron microscope, and computationally reconstructed into 3D density volumes. A major computational challenge is \textit{ab initio} reconstruction, i.e., recovering 3D structure(s) without any image poses or initial volumes. Over the past decade, methods and software for \textit{ab initio} 3D reconstruction and refinement have matured, enabling routine structure determination for well-behaved, structurally homogeneous (i.e. rigid) biomolecules~\cite{scheres2012relion,punjani2017cryosparc,lyumkis2013likelihood,grant2018cis}. Even for more dynamic biomolecular complexes such as macromolecular machines, standard pipelines are often sufficient to align particle images into a common reference frame, thus enabling downstream advanced heterogeneity analysis tools to recover rich conformational ensembles~\cite{zhong2021cryodrgn,punjani20213dva,nakane2018characterisation,donnat2022deep}. 

Modern cryo-EM experiments, however, increasingly image less-purified samples containing a mixture of distinct molecular species~\cite{nogales2024bridging,ho2020bottom}. When multiple species are present, there is no common reference frame, and image alignment can no longer be decoupled from resolving the heterogeneity. Poses, per-particle class assignments, and multiple volumes must instead be inferred jointly, a substantially harder optimization. Multiclass \textit{ab initio} reconstruction is the standard approach to this problem, partitioning particle images into a fixed number of classes and reconstructing one volume per class~\cite{punjani2017cryosparc}. Recent advances, including neural field approaches such as cryoDRGN-AI~\cite{zhong2021cryodrgn} and neural mixture models such as Hydra~\cite{levy2024mixture}, have proposed more expressive models of heterogeneity that are jointly optimized with image poses for heterogeneous \textit{ab initio} reconstruction. Despite this progress, current methods fail when the number of distinct species grows large, as this poses a much more challenging optimization problem~\cite{jeon2024cryobench}. In practice, practitioners compensate with ad hoc strategies: running reconstructions repeatedly with different random seeds, manually inspecting and filtering the results, and iterating until satisfactory classes emerge. These workflows are labor-intensive, difficult to reproduce, and scale poorly with mixture complexity.

Here, we observe that multiclass \textit{ab initio} reconstruction jobs, despite failing to resolve a complex mixture in full, carry weak classification signal. This is analogous to the classical weak learning setting~\cite{schapire1990strength}, where learners that perform only slightly better than chance can be aggregated into a strong classifier. Building on this connection and related work on cluster ensembles~\cite{strehl2002cluster, fred2005combining}, we introduce \textit{reconstruction ensembles}, a meta-algorithm for systematically generating, weighting, and aggregating \textit{ab initio} reconstruction jobs to resolve complex mixtures that no single run can handle. Using cryoSPARC's multiclass reconstruction methods~\cite{punjani2017cryosparc} as weak classifiers, we demonstrate state-of-the-art \textit{ab initio} reconstruction of synthetic and real datasets, including successful recovery of complex mixtures from datasets containing more than tens of species for the first time. The approach is tool-agnostic, requires no modifications to existing reconstruction software, and lays the groundwork for reproducible, automated cryo-EM workflows. In summary, we make the following contributions:

\begin{itemize}
    \item A meta-algorithm for spawning and aggregating \textit{ab initio} reconstruction algorithms to produce final particle classification assignments,
    \item State-of-the-art \textit{ab initio} reconstruction of datasets containing complex mixtures, achieving 97\% classification accuracy of a 45-class subset of Tomotwin-100 and 75\% accuracy on the full Tomotwin-100 dataset~\cite{jeon2024cryobench},
    \item A workflow automation framework built on the cryoSPARC-tools API that enables reproducible, programmatic control of cryo-EM reconstruction pipelines.
\end{itemize}

\section{Background and related work}
 
\subsection{Single particle cryo-EM reconstruction}
In single particle cryo-EM, an aqeuous solution of molecules is vitrified and imaged in a transmission electron microscope, producing $10^{4}$--$10^{7}$ noisy 2D projection images. Each image $X_i$ captures a separate copy of the molecule at an unknown orientation, and can be modeled as~\cite{levy2024mixture, vulovic2013image}:
\begin{equation}
    X_i = C_i \cdot \mathcal{P}_{\phi_i} V_i + \eta_i
\end{equation}
where $V_i : \mathbb{R}^3 \to \mathbb{R}$ is the 3D electron scattering potential (density map), $\phi_i = (R_i, \mathbf{t}_i)$ is the unknown pose consisting of a rotation $R_i \in SO(3)$ and in-plane translation $\mathbf{t}_i \in \mathbb{R}^2$, $\mathcal{P}_{\phi_i}$ is the projection operator that integrates $V_i$ along the imaging axis at orientation $\phi_i$, $C_i$ is the contrast transfer function (CTF) of the microscope, and $\eta_i \sim \mathcal{N}(0, \sigma^2)$ is additive noise. The images are recorded on a discrete $D \times D$ grid with typical signal-to-noise ratios between $10^{-1}$ and $10^{-2}$.
 
The central computational task is to recover $V$ from these projections, which requires jointly solving for the unknown volume and the per-image poses. Expectation-maximization and coordinate ascent algorithms are typically used to find a maximum \textit{a posteriori} estimate of $V$, marginalizing over the pose posterior for each image~\cite{scheres2012relion}. This iterative refinement is sensitive to the initial volume estimate; stochastic gradient descent is commonly used for \textit{ab initio} reconstruction to provide a starting model from scratch~\cite{punjani2017cryosparc}.
 
\subsection{Multiclass \textit{ab initio} reconstruction}
Real cryo-EM samples often contain mixtures of distinct molecular species or conformational states. The standard approach to this \textit{heterogeneity problem} extends the image formation model to assume that images are generated from $K$ independent volumes $V_1, \dots, V_K$. Inference then requires marginalization over both the per-image poses and class assignment probabilities $\pi_j$:
\begin{equation}
    \arg\max_{V_1, \dots, V_K} \sum_{i=1}^{N} \log \sum_{j=1}^{K} \left[ \pi_j \int p(X_i \mid \phi, V_j) \, p(\phi) \, d\phi \right] + \sum_{j=1}^{K} \log p(V_j)
    \label{eq:multiclass}
\end{equation}
This formulation, commonly called multiclass refinement or 3D classification, is implemented in major software packages such as RELION~\cite{scheres2012relion} and cryoSPARC~\cite{punjani2017cryosparc}. In the \textit{ab initio} setting, the algorithm must simultaneously infer both the image class assignments and poses from scratch (i.e. random initialization), making the optimization landscape considerably more challenging than the refinement or classification case where initial volume models or fixed poses are provided, respectively.
 
A key limitation of multiclass \textit{ab initio} reconstruction is that the number of classes $K$ must be specified in advance and is typically kept small (e.g., $K \leq 10$). When the true number of species $M$ exceeds $K$, the algorithm produces a coarse partition that merges distinct species into shared classes. When $K$ is set too large, the optimization frequently fails to converge or produces degenerate solutions. In practice, this means that for complex mixtures containing tens of distinct species, no single \textit{ab initio} run can resolve the full composition of the sample.
 
\subsection{Heterogeneity analysis beyond discrete classes}
A substantial body of work has been devoted to expanding the types of heterogeneity that cryo-EM methods can capture~\cite{donnat2022deep, jeon2024cryobench}. While early methods modeled heterogeneity as a discrete mixture of independent, homogeneous volumes~\cite{scheres2012relion, punjani2017cryosparc}, more recent approaches have moved toward continuous representations of structural variability, including linear subspaces~\cite{punjani20213dva}, neural fields~\cite{zhong2021cryodrgn, zhong2021cryodrgn2}, Gaussian mixture models~\cite{chen2021deep}, and parametric deformation fields~\cite{punjani20233dflex}. However, the majority of these methods operate with fixed poses obtained from an upstream consensus reconstruction, limiting their applicability when the sample contains strong compositional heterogeneity that prevents reliable pose estimation in the first place.
 
In the \textit{ab initio} setting where poses must be inferred jointly with the heterogeneity, fewer methods are available. The multiclass \textit{ab initio} capabilities in RELION and cryoSPARC (Eq.~\ref{eq:multiclass}) model discrete mixtures but are limited to small $K$. CryoDRGN-AI~\cite{levy2025cryodrgnai} demonstrated a fully autodecoding framework for \textit{ab initio} heterogeneous reconstruction, and Hydra~\cite{levy2024mixture} extended this with a mixture of $K$ neural fields to jointly capture compositional and conformational heterogeneity. Stepping beyond the capabilities of a single call to these methods, Lauzirika et al.~\cite{lauzirika2025many} proposed a statistical framework for exploring the number of discrete classes using assignment consistencies across replicate runs, an intuition closely related to ours, which we extend through a full ensemble aggregation framework. Despite this progress, resolving complex mixtures containing tens or more distinct species at the \textit{ab initio} stage remains an open challenge, with no existing methods, to our knowledge, able to recover any of the underlying structures from the Tomotwin-100 benchmark dataset~\cite{jeon2024cryobench}. 
 
\subsection{Workflow automation in cryo-EM}
The cryo-EM data processing pipeline has historically been driven by expert users interacting with graphical interfaces. Recently, several efforts have begun to automate portions of this workflow. The cryoSPARC team demonstrated an end-to-end Workflow feature~\cite{cryosparc2025end}, and the CryoWizard pipeline~\cite{yan2025cryoief} automates particle curation using the Cryo-IEF model. CryoSift~\cite{schafer2025cryosift} provides another example of programmatic interaction with the cryoSPARC suite. Our ensemble framework builds directly on this trend, using the cryoSPARC-tools API to orchestrate the generation and aggregation of reconstruction jobs without manual intervention.
 
\subsection{Weak learning and cluster ensembles}
A central result in machine learning is that weak learners, i.e.~classifiers performing only slightly better than random chance, can be aggregated into strong learners with substantially better performance guarantees~\cite{schapire1990strength}. This observation has driven decades of research on boosting and ensemble methods, producing canonical algorithms such as AdaBoost~\cite{freund1997adaboost} and XGBoost~\cite{chen2016xgboost}.
 
In the unsupervised setting, where labeled feedback is unavailable, a related subfield of \textit{cluster ensembles} has emerged~\cite{strehl2002cluster}. Given multiple clustering assignments of the same dataset, the goal is to aggregate them into a single, stronger partition. A canonical approach due to Fred and Jain~\cite{fred2005combining} constructs a co-assignment matrix recording how often pairs of data points are placed in the same cluster across ensemble members, then applies hierarchical agglomerative clustering to this accumulated evidence. Alternative formulations include graph partitioning~\cite{fern2004graph} and information-theoretic objectives~\cite{strehl2002cluster}. Our weighted similarity score builds directly on the co-assignment matrix framework of~\cite{fred2005combining}, extending it with coverage-aware masking and interpretable weight heuristics tailored to the structure of cryo-EM reconstruction outputs.

\section{Methods}
\begin{wrapfigure}[21]{R}{0.55\textwidth}
    \includegraphics[width=0.53\columnwidth]{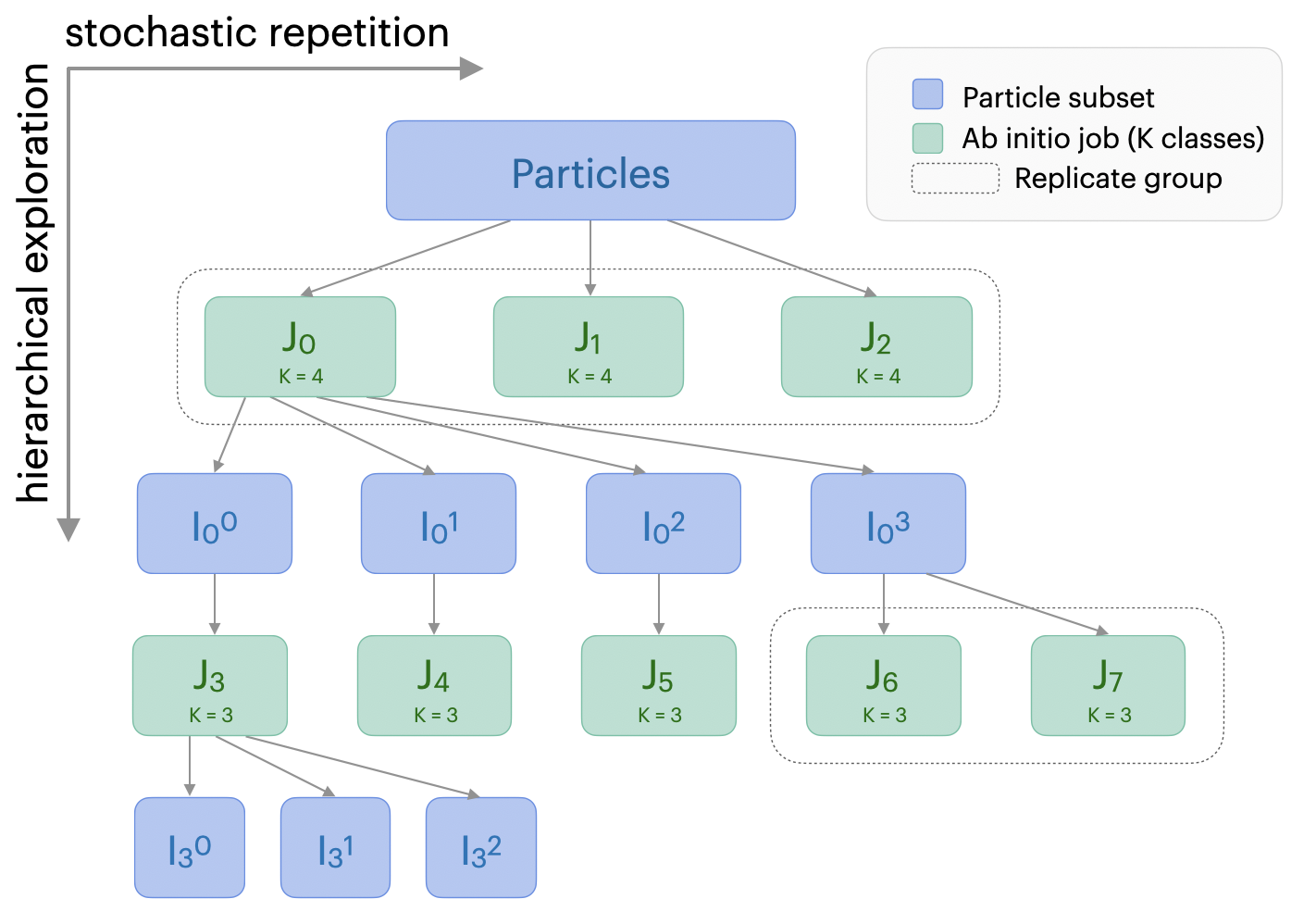}
    \caption{
      Overview of \textit{ab initio} ensemble generation. Along the horizontal axis, \textbf{stochastic repetition} produces replicate jobs that share the same input subset $I_m$ and class count $K_m$ but differ in random seed $\rho_m$. Along the vertical axis, \textbf{hierarchical exploration} launches new jobs on the children $I_m^k$ of a parent job.
    }
    \label{fig:ensemble}
\end{wrapfigure}

Our algorithm proceeds in three stages: ensemble generation, weighted similarity scoring, and clustering. We first formalize the tool abstraction and the two primitives for building ensembles (Fig.~\ref{fig:ensemble}), then describe the aggregation framework that produces the final particle assignments (Fig.~\ref{fig:similarity}).
 
\subsection{Notation and tool abstraction}
We assume access to a stochastic multiclass \textit{ab initio} reconstruction tool that takes a set of $N$ particle images and assigns each particle to one of $K$ classes. Let $I = \{0, \dots, N-1\}$ denote the particle index set.
 
We index the jobs in the ensemble with a single flat counter: $J_0, J_1, \dots, J_{T-1}$ for $T$ total jobs. Each job $J_m$ is characterized by its input particle subset $I_m \subseteq I$, its class count $K_m$, and its random seed $\rho_m$. Annotating the set of possible class labels for a particle to be $[K_m] = \{0, 1, \dots, K_{m}-1\}$, the tool returns a label vector $l_m$ over the input particles:
\begin{equation}
    J_m (I_m; K_m, \rho_m) = \texttt{AbInitio}(I_m; K_m, \rho_m) = l_m \in [K_m]^{|I_m|}
\end{equation}
The label vector partitions $I_m$ into disjoint subsets $I_m^k = \{i \in I_m \mid l_m(i) = k\}$ for $k \in [K_m]$, which we call the \textbf{children} of $J_m$. If the tool produces soft class probabilities rather than hard assignments, we take the argmax as the label. Any other tool-specific parameter is held fixed across the ensemble.

The particles may originate from $M \geq 1$ true underlying species. Our only assumption is that the tool assigns particles better than uniform random, regardless of the relationship between $K_m$ and $M$. This is sufficient for the tool to serve as a weak classifier.

\subsubsection{Ensemble generation}
The ensemble $\mathcal{J} = \{J_0, \dots, J_{T-1}\}$ is built from two primitives, visually summarized in Figure~\ref{fig:ensemble}.
 
\textbf{Stochastic repetition.} Given a particle subset $I_m$ and a class count $K_m$, this primitive launches $R$ replicate jobs with identical parameters while varying the random seed $\rho_m$. Cross-replicate consistency in assignments serves as a proxy for classification confidence.
 
\textbf{Hierarchical exploration.} Given a completed job $J_m$, this primitive launches a new job on each of its children $I_m^0, \dots, I_m^{K_m - 1}$, with a possibly different class count $K'$. This divide-and-conquer strategy enables the ensemble to probe a number of true classes that grows exponentially in the tree depth.
 
These two primitives can be composed freely. The similarity score aggregation (described below) accommodates any job tree or forest, so no particular generation strategy is required.
 
\subsubsection{Weighted similarity score}
\begin{figure}[t]
    \includegraphics[width=1.0\columnwidth, trim={0 4cm 0 0}, clip]{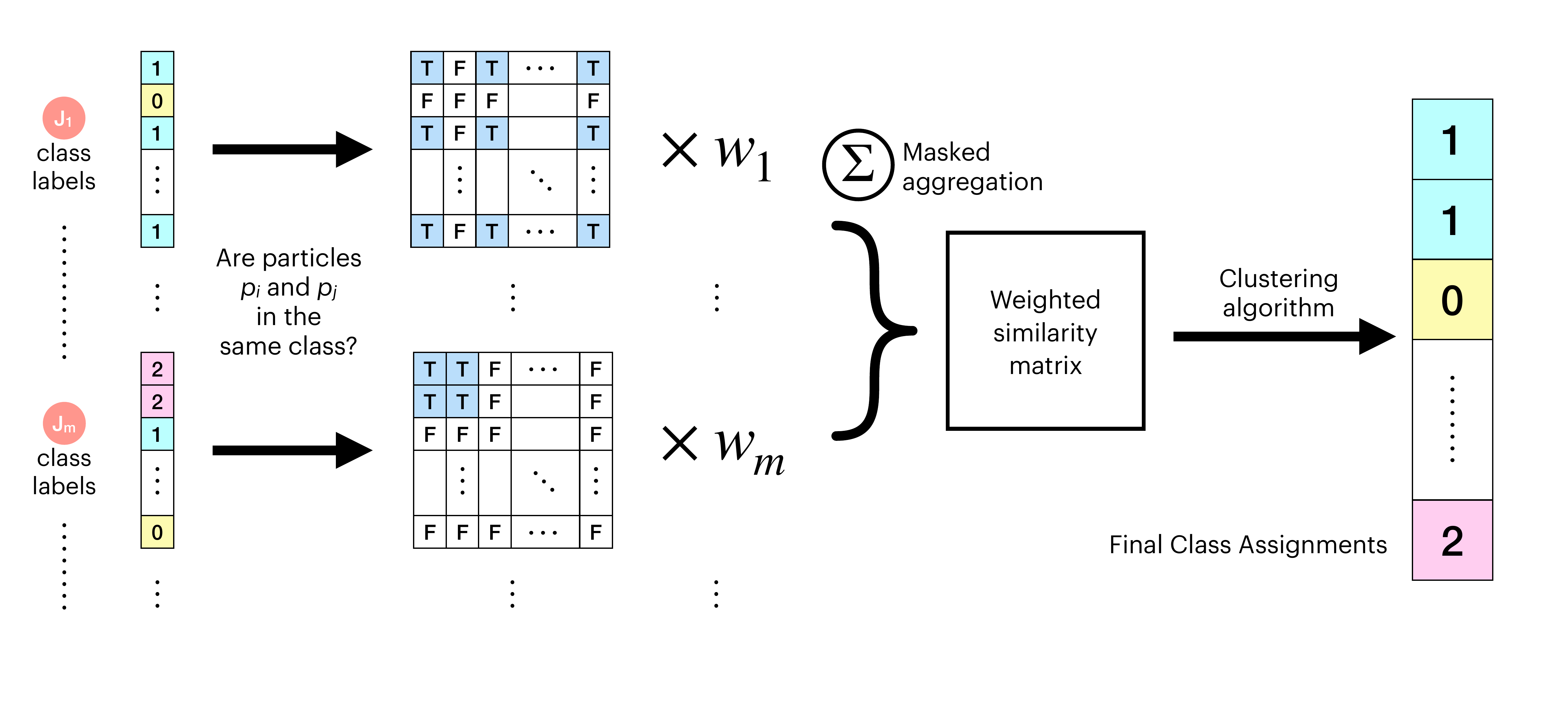}
    \caption{
      Overview of the job aggregation framework that produces final particle assignments. For every job $J_m$ in the ensemble $\mathcal{J}$, the particle co-assignment matrix is calculated, then averaged with weights $w_m$ to yield a similarity score matrix. The similarity score matrix is fed to any off-the-shelf clustering algorithm that accepts pairwise distance inputs to yield final class assignments.
    }
    \label{fig:similarity}
\end{figure}

Given the completed ensemble $\mathcal{J} = \{J_0, \dots, J_{T-1}\}$, we aggregate per-job class assignments into a single inter-particle similarity matrix $S$ (Fig.~\ref{fig:similarity}). This proceeds in two stages: computing per-job co-assignment matrices, then taking their weighted average.
 
For each job $J_m$ with input subset $I_m \subseteq I$, the \textbf{co-assignment matrix} $C^m \in \{0, 1\}^{N \times N}$ records whether two particles $(i, j)$ received the same class label, i.e. $C^m_{ij} = \mathbb{I}\{ l_m(i) = l_m(j) \}$. Furthermore, a \textbf{coverage mask} $\Omega^m$ tracks which particle pairs were present in the job, masking out any pair that was not present in both, i.e. $\Omega^m_{ij} = \mathbb{I}\{ i \in I_m \} \cdot \mathbb{I}\{ j \in I_m \}$.

The \textbf{similarity} between particles $i$ and $j$ is then the weighted mean of their co-assignment values across all jobs in the ensemble that covered both, i.e. their \textbf{coverage-aware weighted average}:
\begin{equation}
    S_{ij} \coloneqq \frac{\sum_{m=0}^{T-1} w_m \, \Omega^m_{ij} \, C^m_{ij} }{\sum_{m=0}^{T-1} w_m \, \Omega^m_{ij}} \in [0, 1]
\end{equation}

The weights in this aggregation routine can be supplied by the expert user if desired. However, instead of requiring the user to manually inspect and assign per-job weights, we also provide \textbf{weight heuristics} to automatically compute $w_m$ as a product of three interpretable penalty factors: (1) jobs that underuse their allotted classes $(w^{\text{eff}}_m)$, (2) jobs that fragment particles into tiny clusters $(w^{\text{tiny}}_m)$, and (3) jobs that funnel most particles into a single dominant class $(w^\text{{sink}}_m)$. All three are derived from the class proportions of job $J_m$. We write $p_m^k = n_m^k / N_m$ for the fraction of the $N_m = |I_m|$ input particles assigned to class $k$, where $\sum_k n_m^k = N_m$.
\begin{equation}
    w_m \coloneqq w^{\text{eff}}_m \cdot w^{\text{tiny}}_m \cdot w^\text{{sink}}_m
\end{equation}
 
The \textbf{effective-class weight} $w^{\text{eff}}_m$ penalizes jobs that concentrate particles into fewer classes than requested. We measure the effective number of occupied classes via the inverse Simpson index~\cite{simpson1949measurement} (equivalently, the inverse Herfindahl-Hirschman index~\cite{rhoades1993herfindahl}), $K_m^{\text{eff}}$. The penalty is Gaussian in log-space, so that $w^{\text{eff}}_m = 1$ when classes are perfectly balanced ($K_m^{\text{eff}} = K_m$) and decays smoothly with bandwidth $\sigma$ as the distribution becomes more skewed:
\begin{equation}
    w_m^{\text{eff}} = \text{exp}\left(-\frac{(\log{K_m^{\text{eff}}} - \log{K_m})^2}{2\sigma^2}\right), \quad \text{here} \quad K_m^{\text{eff}} = \frac{1}{\sum_k (p_{m}^k)^2}
\end{equation}
 
The \textbf{tiny-class weight} $w^{\text{tiny}}_m$ penalizes jobs that produce classes with less than $p^{\text{min}}$ fraction of particles, which are too small for reliable downstream reconstruction. The total probability mass in such classes is $t_{m} = \sum_{k} \mathbb{I}\{p_m^k < p^{\text{min}}\}\; p_m^k$, and the penalty decays exponentially on $t_m$ with rate $\gamma$. The \textbf{sink-class weight} $w^{\text{sink}}_m$ penalizes jobs in which one dominant class absorbs most particles, since such a job provides little discriminative signal to the ensemble. Writing $p_m^{\text{max}} = \max_{k} p_m^k$, the penalty decays exponentially on $p_m^{\text{max}}$ with rate $\beta$.
\begin{equation}
    w_m^{\text{tiny}} = \exp\left(-\gamma \, t_{m}\right), \quad w_m^{\text{sink}} = \exp(-\beta \, p_m^{\max})
\end{equation}
Alternative functional forms (e.g., hard thresholds) are compatible with the framework, where custom weight heuristics could be engineered at will. For our purposes, the exponential form sufficed while providing simplicity in interpretation.
 
\subsubsection{Clustering}
Once $S$ is computed, any clustering algorithm that accepts pairwise similarities (or distances via $1-S$) can be applied. We use agglomerative clustering with average linkage, as it allows straightforward exploration of how assignments change with the number of clusters.
 
A practical issue arises from the dense structure of $S$: with average linkage, large groups of weakly similar particles can dominate the merge criterion over smaller groups of strongly similar particles, causing the algorithm to absorb rare species into majority clusters prematurely. To address this, we apply a \textbf{sparsification} step before clustering. For each row of $S$, we retain only the top $n_{\text{nbr}}$ strongest similarity scores and zero out the rest, effectively constructing a nearest-neighbor graph. We also zero out the diagonal (trivial self-similarity) and symmetrize the result by taking $S_{ij} \leftarrow \max(S_{ij}, S_{ji})$. The final cluster assignments are obtained by cutting the agglomerative dendrogram at the desired number of classes.
 
\section{Results}

We first characterize the performance of the individual \textit{ab initio} reconstruction jobs on datasets of increasing heterogeneity. We then evaluate our ensembling approach on \textit{ab initio} reconstruction of synthetic datasets containing up to 100 distinct biomolecular complexes as well as an experimental cryo-EM dataset. Finally, we analyze how performance scales as a function of ensemble growth and perform hyperparameter ablations of the aggregation step.

\subsection{The \textit{ab initio} reconstruction tool}

Although our approach is tool-agnostic for any multiclass reconstruction algorithm, we use the ``Ab-Initio Reconstruction'' job in cryoSPARC software suite~\cite{punjani2017cryosparc}, which provides programmatic access and job management through a Python API exposed in the \href{https://github.com/cryoem-uoft/cryosparc-tools}{cryosparc-tools} library.

We use only the \textit{default} parameters in version 4.7.1, except the number of classes $K$ and the random seed $\rho$. Given the extensive tunability of this job, we note that our results could be likely be further improved by optimizing the parameters, e.g.~the Fourier radius step in smaller molecules~\cite{kim2025hrhair}. Here, we focus on highlighting the capability of ensembling even with default job settings.

\subsubsection{Datasets}
The Tomotwin-100 dataset, proposed by the CryoBench benchmark suite~\cite{jeon2024cryobench}, consists of 100 structures of various molecular weights and symmetries, with 1,000 images each $(D=128)$, uniformly distributed across the viewing sphere. We focus our analysis on this dataset, since to the best of our knowledge, the complete \textit{ab initio} reconstruction of this dataset has not been reported yet, where the current state-of-the-art analysis provided in Hydra~\cite{levy2024mixture} only focused on a 3-class subset of this 100-class dataset.

We also analyze a 45-class subset of the asymmetric complexes in Tomotwin-100, which we call Tomotwin-45-asym. Evaluating pose accuracy on the full dataset is complicated by symmetry: when the underlying structure possesses rotational symmetry, an image can be explained by multiple equivalent poses, making per-image pose error ill-defined~\cite{levy2024mixture,hodavn2016evaluation}. To avoid complexity from symmetry-aware pose error functions~\cite{hodavn2016evaluation}, we restrict our pose evaluation to structures with C1 symmetry in Tomotwin-100, yielding 45 classes and $N=45,000$ particles.

To test our method beyond synthetic data, we use EMPIAR-10076~\cite{empiar10076}, a standard benchmark for heterogeneous reconstruction of compositional heterogeneity~\cite{empiar10076,zhong2021cryodrgn} consisting of \textit{E.~coli} large ribosomal subunit assembly intermediates ($N=131,899$ images, $D=320$). The dataset contains four major structural states, two rare species, and a substantial junk fraction. We process the full, unfiltered stack in our reconstruction.

To compute classification accuracy, in order to assess the correct permutation of labels, we follow a simple majority vote within every cluster to determine which ground truth label it maps to. In pose accuracy measurements, the final class assignments are used to generate per-class homogeneous reconstructions, which are then aligned with the mapped ground truth poses with a global rigid-body alignment.  
\subsection{Performance of the one-shot ab initio jobs}

\begin{figure} [h]
  \begin{center}
    \includegraphics[width=0.9\columnwidth, trim={0 0.5cm 0 0cm}, clip]{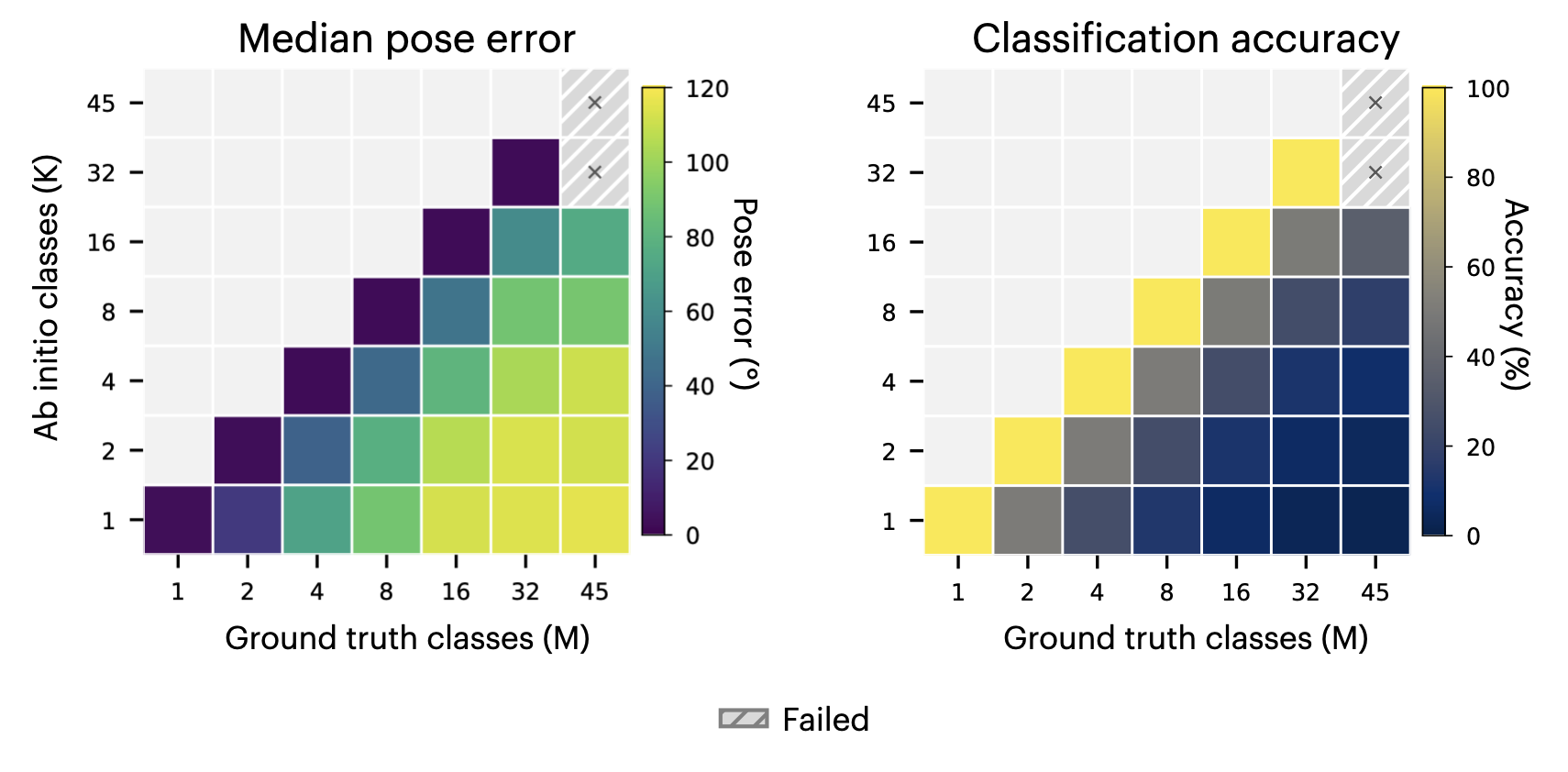}
    \caption{
      Median angular error (left) and classification accuracy (right) of single-shot ab initio jobs on subsets of Tomotwin-45-asym with increasing heterogeneity.
    }
    \label{fig:oneshot_hmap}
  \end{center}
\end{figure}

We first run the cryoSPARC~\cite{punjani2017cryosparc} ``Ab-Initio Reconstruction'' job on subsets of Tomotwin-45-asym with increasing number of ground truth classes ($M = 1, 2, \dots, 45$) and varying the class number parameter $K$. This sweep serves two purposes: it establishes a baseline for single-shot reconstruction performance, and it validates the coarse clustering assumption ($K < M$) that our ensemble design relies on. The median angular pose error and classification accuracy are reported in Figure~\ref{fig:oneshot_hmap}, with the complete suite of metrics in Appendix~\ref{app:sshot}.

On the diagonal ($K = M$), the job achieves near-perfect classification and pose estimation up to $M = 32$, with median pose errors of 2--3 degrees. As the ground truth complexity grows relative to the job's class budget (moving rightward or downward from the diagonal) both accuracy and pose quality degrade gracefully. On the full dataset ($M = 45$), the job fails to converge at large $K$ values ($K \geq 32$) but still converges for coarser settings ($K \leq 16$), confirming that it can serve as a weak classifier even when it cannot resolve the full mixture.

\subsection{Performance of the ab initio ensemble}

\begin{table}[h]
  \caption{Heterogeneous ab initio reconstruction performance measured by classification, clustering, and pose metrics for the Tomotwin benchmark datasets. For the baseline one-shot strategy, the best performing jobs are shown, i.e.~$K=16$ for both datasets.}
  \label{tab:results}
  \begin{center}
    \begin{small}
      \begin{sc}
        \begin{tabular}{llccc cc}
          \toprule
          Dataset & Strategy
          & \multicolumn{3}{c}{Classification (\%)}
          & \multicolumn{2}{c}{Pose Error ($\deg$)} \\
          \cmidrule(lr){3-5} \cmidrule(lr){6-7}
          & 
          & Accuracy & Precision & Macro F1
          & Mean & Median \\
          \midrule
          \multirow{2}{*}{TT-45-asym}
          & cryoDRGN2 & 13.1 & 10.1 & 10.9 & 124.6 & 128.3 \\
          & cryoDRGN-AI & 9.3 & 6.8 & 7.4 & 124.2 & 129.3 \\
          & Hydra ($K=4$) & 4.4 & 0.4 & 0.7 & 124.4 & 129.4 \\
          & One-Shot & 34.6 & 25.4 & 22.7 & 78.3 & 74.1 \\
          & Ensemble   & \textbf{97.4} & \textbf{97.4} & \textbf{97.4} & \textbf{6.11} & \textbf{1.71} \\
          \midrule
          \multirow{2}{*}{TT-100}
          & One-Shot   & 15.2 & 3.5 & 5.5 & - & - \\
          & Ensemble   & \textbf{75.0} & \textbf{71.3} & \textbf{71.9} & - & - \\
          \bottomrule
        \end{tabular}
      \end{sc}
    \end{small}
  \end{center}
  \vskip -0.1in
\end{table}

We evaluate our ensemble method on two datasets of increasing difficulty: the 45-class Tomotwin-45-asym and the full 100-class Tomotwin-100. For both datasets, the ensemble was generated using $K=\{ 4, 8, 16 \}$ job trees at differing $2-3$ levels of hierarchical depth with $3-5$ replicates per level (Appendix~\ref{app:mainresults}). For Tomotwin-45-asym, its ensemble totaled $257$ jobs and $332$ H100-hours with a critical path of $4.3$ wall-hours; while for Tomotwin-100, its ensemble totaled $235$ jobs and $320$ hours with a critical path of $4.3$ wall-hours. In both cases, default parameters were used for all jobs. Further experimental details are provided in Appendix~\ref{app:mainresults}.
 
We compare against four baselines: the best single-shot cryoSPARC run ($K=16$ for both datasets), and three heterogeneous reconstruction methods: cryoDRGN2~\cite{zhong2021cryodrgn2}, cryoDRGN-AI~\cite{levy2025cryodrgnai}, and Hydra ($K=4$)~\cite{levy2024mixture}. These three methods were run with default settings (Appendix~\ref{app:mainresults}). Because these methods were designed for smaller mixtures, we include them primarily to illustrate the difficulty of the 45-class setting rather than as matched comparisons. Results are summarized in Table~\ref{tab:results}.

On Tomotwin-45-asym, the ensemble achieves 97.4\% classification accuracy and a median pose error of 1.71\textdegree{}, compared to 34.6\% accuracy and 74.1\textdegree{} median pose error for the best single-shot run. The three neural-field baselines produce pose errors near the uniform-random baseline of ${\sim}$130\textdegree{}, indicating that these methods cannot resolve the mixture at this complexity. The ensemble method fully reconstructs this 45-class mixture, as seen from the volume renderings in Appendix~\ref{app:tt45maps}. Furthermore, on Tomotwin-100, the ensemble reaches 75.0\% accuracy -- a five-fold improvement over the 15.2\% one-shot baseline. To our knowledge, neither the 45-class nor the 100-class Tomotwin benchmarks have been previously resolved at the \textit{ab initio} stage, the prior state-of-the-art analysis in Hydra~\cite{levy2024mixture} was limited to a 3-class subset. Our results therefore represent an order-of-magnitude expansion in the mixture complexity that \textit{ab initio} methods can handle.

\subsection{Scaling analysis in ensemble growth}

\begin{figure}[h]
  \begin{center}
    \includegraphics[width=1.0\columnwidth, trim={0 0cm 0 4cm}, clip]{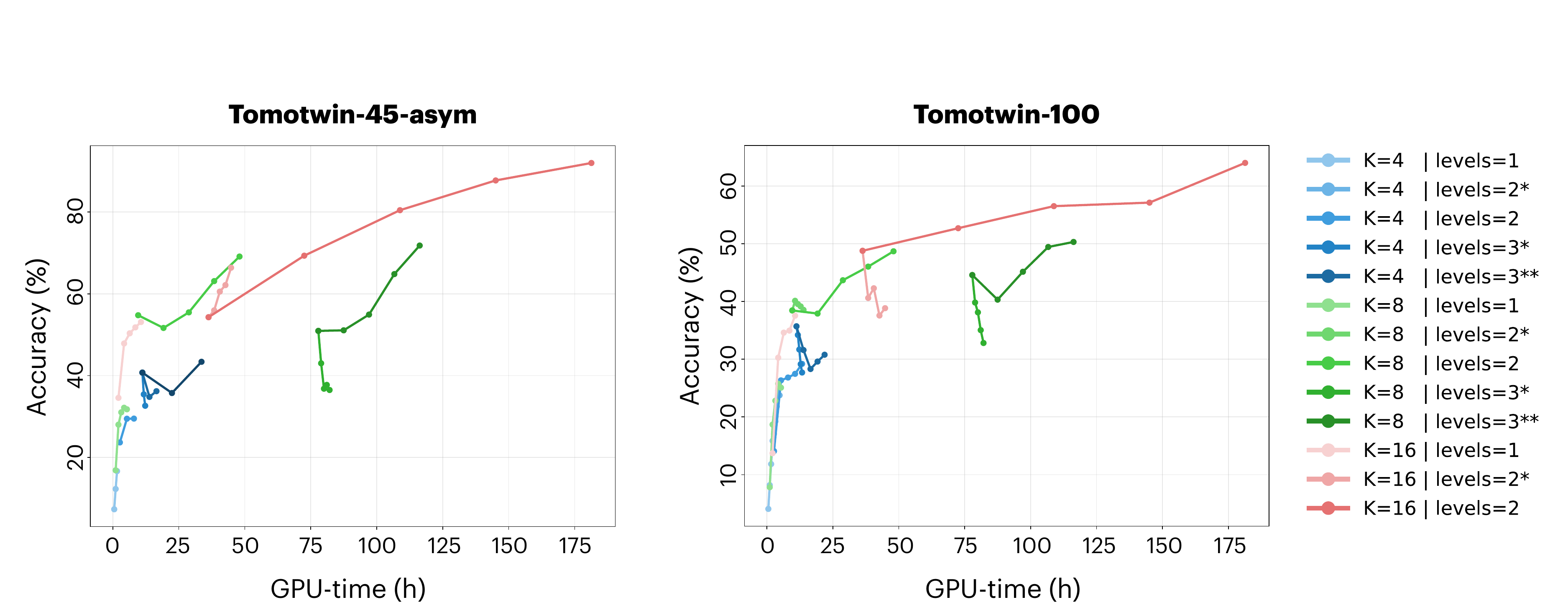}
    \caption{
      Utility-cost frontiers of scaling job ensembling for Tomotwin-45-asym (left) and Tomotwin-100 (right). The vertical (utility) axis is classification accuracy, and the horizontal (cost) axis is the number of H100 GPU-hours used. Each curve shows ensembling up to 5 replicates and different curves indicate different hierachical levels with a fixed class $K$.
    }
    \label{fig:scaling}
  \end{center}
\end{figure}

To understand how classification accuracy scales with compute, we systematically vary both the depth of the job tree (hierarchical exploration) and the number of random replicates (stochastic repetition), sweeping over $K \in \{4, 8, 16\}$. To simplify the configuration space, each ensemble uses a single $K$ value across all jobs; the reported accuracies are therefore lower bounds on what mixed-$K$ ensembles could achieve. Figure~\ref{fig:scaling} plots the resulting accuracy-compute trade-off surfaces for Tomotwin-45-asym and Tomotwin-100, with further details in Appendix~\ref{app:scaling}.

On both datasets, accuracy typically improves with the first few levels of hierarchical depth and with additional replicates. Because replicates within a level run in parallel, the wall-clock time scales with tree depth rather than total job count, growing only logarithmically in the ensemble size. The ``level'' annotations in Figure~\ref{fig:scaling} indicate the depth of the deepest job tree in each ensemble.

\begin{wrapfigure}[15]{R}{0.48\textwidth}
    \includegraphics[width=0.45\columnwidth]{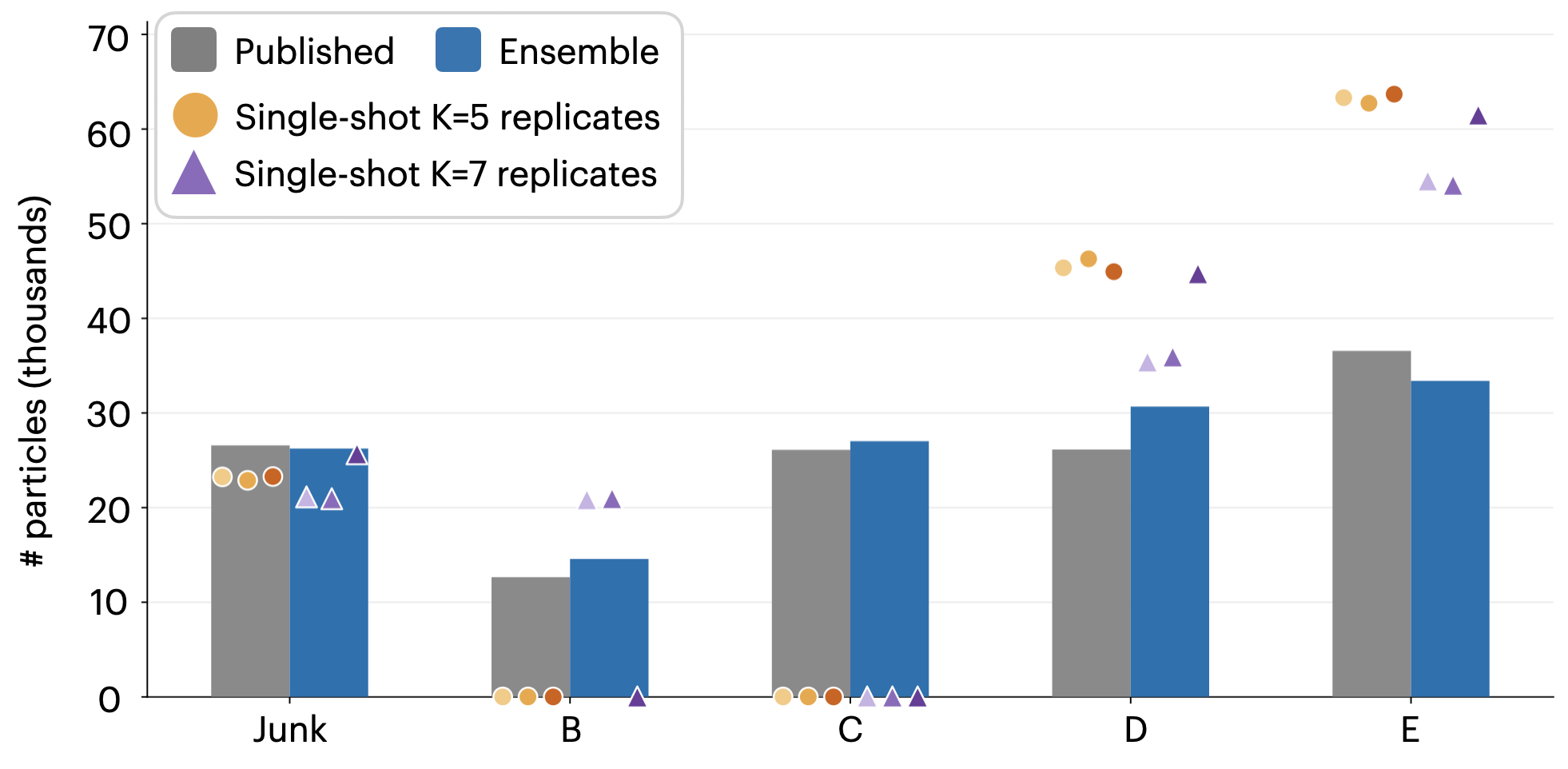}
    \caption{
      Particle counts of our reconstruction ensembling approach, the published major labels~\cite{empiar10076}, and single-job baselines (each repeated with 3 different seeds).
    }
    \label{fig:bar50s}
\end{wrapfigure}

\subsection{Real dataset}

We apply our ensemble method to EMPIAR-10076~\cite{empiar10076}, processing the full unfiltered particle stack ($N = 131{,}899$) without removing junk particles. The ensemble was generated using $3$ replicates of $\{4, 8, 16\}$-class jobs at $2$ levels of hierarchical depth, and $6$-class jobs at $3$ levels of hierarchical depth, totaling $222$ jobs and $250$ H100-hours with a critical path of $4.3$ wall-hours. We cut the final dendrogram at 7 classes.

The ensemble successfully separates all four major LSU assembly intermediates (classes B--E) from the junk fraction without manual filtering. Per-class homogeneous refinement yields reconstructions at $\leq 4~\mathring{A}$ resolution for all four states (further experimental details, maps, and FSC curves are in Appendix~\ref{app:50s}). Figure~\ref{fig:bar50s} compares the particle counts in our predicted classes against the published annotations~\cite{empiar10076}, as well as single-shot \textit{ab initio} baselines. Unlike the single-shot baselines which miss out on the B and C major classes, our ensemble class populations are in broad agreement with the published labels, with the ensemble additionally resolving sub-states within classes D and E that are consistent with the finer partitions reported in prior work~\cite{zhong2021cryodrgn, empiar10076}.

\subsection{Hyperparameter ablations}
The aggregation step has five hyperparameters: three exponential rates in the weight heuristics ($\sigma$, $\beta$, $\gamma$), a minimum class size threshold $p^{\text{min}}$, and the sparsification neighborhood size $n_{\text{nbr}}$. We sweep all five on Tomotwin-45-asym (Appendix~\ref{app:hparam}) and find that the weight parameters have minimal impact (fewer than 10 percentage points of variation), while $n_{\text{nbr}}$ is the most consequential choice, producing swings of over 45 percentage points. The main results in Table~\ref{tab:results} use $(\sigma, \beta, \gamma, p^{\text{min}}, n_{\text{nbr}}) = (0.6,\; 6,\; 10,\; 1/90,\; 100)$.

\section{Discussion}
We demonstrate that ensembling \textit{ab initio} reconstruction jobs can resolve complex mixtures in cryo-EM beyond the reach of any single run or any existing method. Because constituent jobs used default settings, we believe these results represent a lower bound on what the framework can achieve with task-specific tuning. By reducing the problem of resolving complex mixtures to the orchestration of fast, coarse reconstruction jobs, the approach opens a path toward high-throughput heterogeneous reconstruction in modern experimental settings where samples contain tens or more distinct species. In doing so, it replaces the ad hoc classification schemes practitioners currently rely on with a systematic procedure that scales with compute rather than manual labor.

The aggregation framework is deliberately flexible. In our experiments, we considered ensembles with a fixed $K$ and a predefined replication strategy for scaling analysis, but the aggregation accepts any collection of multi-class reconstruction jobs regardless of their parameters, input subsets, or tree structure. A practitioner could mix jobs with different $K$ values, incorporate hand-picked runs targeting specific subpopulations, or feed in results from entirely different software packages. This same flexibility makes the framework amenable to autonomous or agent-driven exploration strategies that adaptively allocate compute based on intermediate ensemble results.
 
Several limitations warrant discussion. First, although the aggregation function is principled, it introduces hyperparameters that encode priors on cluster sizes and uniformity. Our ablations show robustness to the weight heuristic parameters, but sensitivity to the sparsification neighborhood size $n_{\text{nbr}}$. Whether our default value for $n_{\text{nbr}}$ generalizes across diverse real datasets remains to be tested. Second, our method classifies particles but does not explicitly model junk; on EMPIAR-10076, junk particles were separated into their own cluster, but our meta-algorithm could benefit from a dedicated filtering step or treatment of a junk or outlier class. Third, the $N \times N$ similarity matrix becomes a memory bottleneck for large particle stacks.\footnote{The required memory was in the tens of GBs for a dataset with $N=131{,}899$ particles.} Sparse storage and graph-based clustering algorithms that avoid materializing the full matrix are natural remedies to enable scaling to datasets with millions of particles. Addressing these limitations would bring the framework closer to routine deployment on the increasingly complex mixtures encountered in modern cryo-EM experiments, including cell lysates, crude extracts, and \textit{in situ} imaging.

\begin{ack}
The authors acknowledge the use of computing resources at Princeton Research Computing, a consortium of groups led by the Princeton Institute for Computational Science and Engineering (PICSciE) and Office of Information Technology’s Research Computing. 
This research has been made possible by in part by grant number 2025-358484 from the Chan Zuckerberg Initiative DAF, an advised fund of Silicon Valley Community Foundation, the National Institutes of Health under grant number DP2GM164606, and the AI2050 program at Schmidt Sciences (Grant G-25-69788).
The Zhong lab is grateful for support from the Princeton Catalysis Initiative, Princeton School of Engineering and Applied Sciences, Janssen Pharmaceuticals, and Generate Biomedicines. The funders had no role in study design, data collection and analysis, decision to publish or preparation of the manuscript.
\end{ack}

\newpage

\bibliography{references}

\begin{thebibliography}{38}
\providecommand{\natexlab}[1]{#1}
\providecommand{\url}[1]{\texttt{#1}}
\expandafter\ifx\csname urlstyle\endcsname\relax
  \providecommand{\doi}[1]{doi: #1}\else
  \providecommand{\doi}{doi: \begingroup \urlstyle{rm}\Url}\fi

\bibitem[Scheres(2012)]{scheres2012relion}
Sjors~HW Scheres.
\newblock {RELION}: implementation of a {B}ayesian approach to cryo-{EM} structure determination.
\newblock \emph{{J}ournal of {S}tructural {B}iology}, 180\penalty0 (3):\penalty0 519--530, 2012.

\bibitem[Punjani et~al.(2017)Punjani, Rubinstein, Fleet, and Brubaker]{punjani2017cryosparc}
Ali Punjani, John~L Rubinstein, David~J Fleet, and Marcus~A Brubaker.
\newblock cryo{SPARC}: algorithms for rapid unsupervised cryo-{EM} structure determination.
\newblock \emph{Nature {M}ethods}, 14\penalty0 (3):\penalty0 290--296, 2017.

\bibitem[Lyumkis et~al.(2013)Lyumkis, Brilot, Theobald, and Grigorieff]{lyumkis2013likelihood}
Dmitry Lyumkis, Axel~F Brilot, Douglas~L Theobald, and Nikolaus Grigorieff.
\newblock {L}ikelihood-based classification of cryo-{EM} images using {FREALIGN}.
\newblock \emph{{J}ournal of {S}tructural {B}iology}, 183\penalty0 (3):\penalty0 377--388, 2013.

\bibitem[Grant et~al.(2018)Grant, Rohou, and Grigorieff]{grant2018cis}
Timothy Grant, Alexis Rohou, and Nikolaus Grigorieff.
\newblock \textit{cis}{TEM}, user-friendly software for single-particle image processing.
\newblock \emph{e{L}ife}, 7:\penalty0 e35383, 2018.

\bibitem[Zhong et~al.(2021{\natexlab{a}})Zhong, Bepler, Berger, and Davis]{zhong2021cryodrgn}
Ellen~D Zhong, Tristan Bepler, Bonnie Berger, and Joseph~H Davis.
\newblock {C}ryo{DRGN}: reconstruction of heterogeneous cryo-{EM} structures using neural networks.
\newblock \emph{{N}ature {M}ethods}, 18\penalty0 (2):\penalty0 176--185, 2021{\natexlab{a}}.

\bibitem[Punjani and Fleet(2021)]{punjani20213dva}
Ali Punjani and David~J Fleet.
\newblock 3{D} variability analysis: {R}esolving continuous flexibility and discrete heterogeneity from single particle cryo-{EM}.
\newblock \emph{{J}ournal of {S}tructural {B}iology}, 213\penalty0 (2):\penalty0 107702, 2021.

\bibitem[Nakane et~al.(2018)Nakane, Kimanius, Lindahl, and Scheres]{nakane2018characterisation}
Takanori Nakane, Dari Kimanius, Erik Lindahl, and Sjors~HW Scheres.
\newblock {C}haracterisation of molecular motions in cryo-{EM} single-particle data by multi-body refinement in {RELION}.
\newblock \emph{e{L}ife}, 7:\penalty0 e36861, 2018.

\bibitem[Donnat et~al.(2022)Donnat, Levy, Poitevin, Zhong, and Miolane]{donnat2022deep}
Claire Donnat, Axel Levy, Frederic Poitevin, Ellen~D Zhong, and Nina Miolane.
\newblock {D}eep generative modeling for volume reconstruction in cryo-electron microscopy.
\newblock \emph{{J}ournal of {S}tructural {B}iology}, 214\penalty0 (4):\penalty0 107920, 2022.

\bibitem[Nogales and Mahamid(2024)]{nogales2024bridging}
Eva Nogales and Julia Mahamid.
\newblock {B}ridging structural and cell biology with cryo-electron microscopy.
\newblock \emph{Nature}, 628\penalty0 (8006):\penalty0 47--56, 2024.

\bibitem[Ho et~al.(2020)Ho, Li, Lai, Terwilliger, Beck, Wohlschlegel, Goldberg, Fitzpatrick, and Zhou]{ho2020bottom}
Chi-Min Ho, Xiaorun Li, Mason Lai, Thomas~C Terwilliger, Josh~R Beck, James Wohlschlegel, Daniel~E Goldberg, Anthony~WP Fitzpatrick, and Z~Hong Zhou.
\newblock {B}ottom-up structural proteomics: cryo{EM} of protein complexes enriched from the cellular milieu.
\newblock \emph{{N}ature {M}ethods}, 17\penalty0 (1):\penalty0 79--85, 2020.

\bibitem[Levy et~al.(2024)Levy, Raghu, Shustin, Peng, Li, Clarke, Wetzstein, and Zhong]{levy2024mixture}
Axel Levy, Rishwanth Raghu, David Shustin, Adele Peng, Huan Li, Oliver Clarke, Gordon Wetzstein, and Ellen Zhong.
\newblock {M}ixture of neural fields for heterogeneous reconstruction in cryo-{EM}.
\newblock \emph{{A}dvances in {N}eural {I}nformation {P}rocessing {S}ystems}, 37:\penalty0 56988--57017, 2024.

\bibitem[Jeon et~al.(2024)Jeon, Raghu, Astore, Woollard, Feathers, Kaz, Hanson, Cossio, and Zhong]{jeon2024cryobench}
Minkyu Jeon, Rishwanth Raghu, Miro Astore, Geoffrey Woollard, Ryan Feathers, Alkin Kaz, Sonya Hanson, Pilar Cossio, and Ellen Zhong.
\newblock {C}ryo{B}ench: {D}iverse and challenging datasets for the heterogeneity problem in cryo-{EM}.
\newblock \emph{{A}dvances in {N}eural {I}nformation {P}rocessing {S}ystems}, 37:\penalty0 89468--89512, 2024.

\bibitem[Schapire(1990)]{schapire1990strength}
Robert~E Schapire.
\newblock {T}he strength of weak learnability.
\newblock \emph{{M}achine {L}earning}, 5\penalty0 (2):\penalty0 197--227, 1990.

\bibitem[Strehl and Ghosh(2002)]{strehl2002cluster}
Alexander Strehl and Joydeep Ghosh.
\newblock {C}luster ensembles---a knowledge reuse framework for combining multiple partitions.
\newblock \emph{{J}ournal of {M}achine {L}earning {R}esearch}, 3\penalty0 (Dec):\penalty0 583--617, 2002.

\bibitem[Fred and Jain(2005)]{fred2005combining}
Ana~LN Fred and Anil~K Jain.
\newblock {C}ombining multiple clusterings using evidence accumulation.
\newblock \emph{{IEEE} {T}ransactions on {P}attern {A}nalysis and {M}achine {I}ntelligence}, 27\penalty0 (6):\penalty0 835--850, 2005.

\bibitem[Vulovi{\'c} et~al.(2013)Vulovi{\'c}, Ravelli, van Vliet, Koster, Lazi{\'c}, L{\"u}cken, Rullg{\aa}rd, {\"O}ktem, and Rieger]{vulovic2013image}
Milo{\v{s}} Vulovi{\'c}, Raimond~BG Ravelli, Lucas~J van Vliet, Abraham~J Koster, Ivan Lazi{\'c}, Uwe L{\"u}cken, Hans Rullg{\aa}rd, Ozan {\"O}ktem, and Bernd Rieger.
\newblock {I}mage formation modeling in cryo-electron microscopy.
\newblock \emph{{J}ournal of {S}tructural {B}iology}, 183\penalty0 (1):\penalty0 19--32, 2013.

\bibitem[Zhong et~al.(2021{\natexlab{b}})Zhong, Lerer, Davis, and Berger]{zhong2021cryodrgn2}
Ellen~D Zhong, Adam Lerer, Joseph~H Davis, and Bonnie Berger.
\newblock {C}ryo{DRGN}2: {A}b initio neural reconstruction of 3{D} protein structures from real cryo-{EM} images.
\newblock In \emph{Proceedings of the IEEE/CVF International Conference on Computer Vision}, pages 4066--4075, 2021{\natexlab{b}}.

\bibitem[Chen and Ludtke(2021)]{chen2021deep}
Muyuan Chen and Steven~J Ludtke.
\newblock {D}eep learning-based mixed-dimensional {G}aussian mixture model for characterizing variability in cryo-{EM}.
\newblock \emph{{N}ature {M}ethods}, 18\penalty0 (8):\penalty0 930--936, 2021.

\bibitem[Punjani and Fleet(2023)]{punjani20233dflex}
Ali Punjani and David~J Fleet.
\newblock 3{DF}lex: determining structure and motion of flexible proteins from cryo-{EM}.
\newblock \emph{Nature Methods}, 20\penalty0 (6):\penalty0 860--870, 2023.

\bibitem[Levy et~al.(2025)Levy, Raghu, Feathers, Grzadkowski, Poitevin, Johnston, Vallese, Clarke, Wetzstein, and Zhong]{levy2025cryodrgnai}
Axel Levy, Rishwanth Raghu, Ryan~J Feathers, Michal Grzadkowski, Fr{\'e}d{\'e}ric Poitevin, Jake~D Johnston, Francesca Vallese, Oliver~Biggs Clarke, Gordon Wetzstein, and Ellen~D Zhong.
\newblock Cryo{DRGN-AI}: neural ab initio reconstruction of challenging cryo-{EM} and cryo-{ET} datasets.
\newblock \emph{Nature Methods}, 22\penalty0 (7):\penalty0 1486--1494, 2025.

\bibitem[Lauzirika et~al.(2025)Lauzirika, Pernica, Herreros, Ram{\'\i}rez-Aportela, Krieger, Gragera, Iceta, Conesa, Fonseca, Jim{\'e}nez, et~al.]{lauzirika2025many}
O~Lauzirika, M~Pernica, D~Herreros, E~Ram{\'\i}rez-Aportela, J~Krieger, M~Gragera, M~Iceta, P~Conesa, Y~Fonseca, J~Jim{\'e}nez, et~al.
\newblock {H}ow many (distinguishable) classes can we identify in single-particle analysis?
\newblock \emph{{A}cta {C}rystallographica {S}ection {D}}, 81\penalty0 (10):\penalty0 535--544, Oct 2025.
\newblock \doi{10.1107/S2059798325007831}.

\bibitem[CryoSPARC et~al.(2025)CryoSPARC, Barber, Bridges, Dawood, Elder, Frasser, Hu, Liu, McLean, Narine, et~al.]{cryosparc2025end}
Team CryoSPARC, Kelly Barber, Hannah Bridges, Suhail Dawood, Katherine Elder, Nick Frasser, Fiona Hu, Serena Liu, Michael McLean, Ryan Narine, et~al.
\newblock {E}nd-to-end automation of repeat-target cryo-{EM} structure determination in {C}ryo{SPARC}.
\newblock \emph{bioRxiv}, 2025.
\newblock \doi{10.1101/2025.10.17.682689}.

\bibitem[Yan et~al.(2026)Yan, Fan, Yuan, and Shen]{yan2025cryoief}
Yang Yan, Shiqi Fan, Fajie Yuan, and Huaizong Shen.
\newblock {A} comprehensive foundation model for cryo-{EM} image processing.
\newblock \emph{{N}ature {M}ethods}, 23:\penalty0 88--95, 2026.

\bibitem[Sch{\"a}fer et~al.(2025)Sch{\"a}fer, Calza, Hom, Damodar, Peng, Bogdanovi{\'c}, Lander, Stagg, and Cianfrocco]{schafer2025cryosift}
J-H Sch{\"a}fer, Austin Calza, Keenan Hom, Puneeth Damodar, Ruizhi Peng, Neboj{\v{s}}a Bogdanovi{\'c}, Gabriel~C Lander, Scott~M Stagg, and Michael~A Cianfrocco.
\newblock {C}ryo{S}ift: an accessible and automated {CNN}-driven tool for cryo-{EM} 2{D} class selection.
\newblock \emph{{S}tructural {B}iology and {C}rystallization {C}ommunications}, 81\penalty0 (12):\penalty0 517--526, 2025.

\bibitem[Freund and Schapire(1997)]{freund1997adaboost}
Yoav Freund and Robert~E Schapire.
\newblock {A} {D}ecision-{T}heoretic {G}eneralization of {O}n-{L}ine {L}earning and an {A}pplication to {B}oosting.
\newblock \emph{{J}ournal of {C}omputer and {S}ystem {S}ciences}, 55\penalty0 (1):\penalty0 119--139, 1997.

\bibitem[Chen and Guestrin(2016)]{chen2016xgboost}
Tianqi Chen and Carlos Guestrin.
\newblock {XGB}oost: {A} {S}calable {T}ree {B}oosting {S}ystem.
\newblock In \emph{{P}roceedings of the 22nd {ACM} {SIGKDD} {I}nternational {C}onference on {K}nowledge {D}iscovery and {D}ata {M}ining}, KDD '16, page 785–794, New York, NY, USA, 2016. Association for Computing Machinery.
\newblock ISBN 9781450342322.
\newblock \doi{10.1145/2939672.2939785}.

\bibitem[Fern and Brodley(2004)]{fern2004graph}
Xiaoli~Zhang Fern and Carla~E Brodley.
\newblock {S}olving cluster ensemble problems by bipartite graph partitioning.
\newblock In \emph{{P}roceedings of the {T}wenty-{F}irst {I}nternational {C}onference on {M}achine {L}earning}, page~36, 2004.

\bibitem[Simpson(1949)]{simpson1949measurement}
Edward~H Simpson.
\newblock {M}easurement of {D}iversity.
\newblock \emph{{N}ature}, 163\penalty0 (4148):\penalty0 688--688, 1949.

\bibitem[Rhoades(1993)]{rhoades1993herfindahl}
Stephen~A Rhoades.
\newblock The {H}erfindahl-{H}irschman {I}ndex.
\newblock \emph{Federal Reserve Bulletin}, 79:\penalty0 188, 1993.

\bibitem[Kim et~al.(2025)Kim, Li, and Clarke]{kim2025hrhair}
Kookjoo Kim, Huan Li, and Oliver~B. Clarke.
\newblock {H}igh-resolution ab initio reconstruction enables cryo-{EM} structure determination of small particles.
\newblock \emph{bioRxiv}, 2025.
\newblock \doi{10.1101/2025.09.08.674935}.

\bibitem[Hoda{\v{n}} et~al.(2016)Hoda{\v{n}}, Matas, and Obdr{\v{z}}{\'a}lek]{hodavn2016evaluation}
Tom{\'a}{\v{s}} Hoda{\v{n}}, Ji{\v{r}}{\'\i} Matas, and {\v{S}}t{\v{e}}p{\'a}n Obdr{\v{z}}{\'a}lek.
\newblock {O}n evaluation of 6{D} object pose estimation.
\newblock In \emph{{E}uropean {C}onference on {C}omputer {V}ision}, pages 606--619. Springer, 2016.

\bibitem[Davis et~al.(2016)Davis, Tan, Carragher, Potter, Lyumkis, and Williamson]{empiar10076}
Joseph~H Davis, Yong~Zi Tan, Bridget Carragher, Clinton~S Potter, Dmitry Lyumkis, and James~R Williamson.
\newblock Modular assembly of the bacterial large ribosomal subunit.
\newblock \emph{Cell}, 167\penalty0 (6):\penalty0 1610--1622, 2016.

\bibitem[Hubert and Arabie(1985)]{hubert1985ari}
Lawrence Hubert and Phipps Arabie.
\newblock {C}omparing partitions.
\newblock \emph{Journal of Classification}, 2\penalty0 (1):\penalty0 193--218, 1985.

\bibitem[Vinh et~al.(2009)Vinh, Epps, and Bailey]{vinh2009ami}
Nguyen~Xuan Vinh, Julien Epps, and James Bailey.
\newblock Information theoretic measures for clusterings comparison: is a correction for chance necessary?
\newblock In \emph{Proceedings of the 26th Annual International Conference on Machine Learning}, ICML `09, page 1073–1080, New York, NY, USA, 2009. Association for Computing Machinery.
\newblock ISBN 9781605585161.
\newblock \doi{10.1145/1553374.1553511}.

\bibitem[Fowlkes and Mallows(1983)]{fowlkes1983fmi}
Edward~B Fowlkes and Colin~L Mallows.
\newblock A method for comparing two hierarchical clusterings.
\newblock \emph{Journal of the American Statistical Association}, 78\penalty0 (383):\penalty0 553--569, 1983.

\bibitem[Arabie and Boorman(1973)]{arabie1973vi}
Phipps Arabie and Scott~A Boorman.
\newblock Multidimensional scaling of measures of distance between partitions.
\newblock \emph{Journal of Mathematical Psychology}, 10\penalty0 (2):\penalty0 148--203, 1973.

\bibitem[Pedregosa et~al.(2011)Pedregosa, Varoquaux, Gramfort, Michel, Thirion, Grisel, Blondel, Prettenhofer, Weiss, Dubourg, Vanderplas, Passos, Cournapeau, Brucher, Perrot, and Duchesnay]{scikit-learn}
F.~Pedregosa, G.~Varoquaux, A.~Gramfort, V.~Michel, B.~Thirion, O.~Grisel, M.~Blondel, P.~Prettenhofer, R.~Weiss, V.~Dubourg, J.~Vanderplas, A.~Passos, D.~Cournapeau, M.~Brucher, M.~Perrot, and E.~Duchesnay.
\newblock Scikit-learn: {M}achine {L}earning in {P}ython.
\newblock \emph{Journal of Machine Learning Research}, 12:\penalty0 2825--2830, 2011.

\bibitem[Pettersen et~al.(2021)Pettersen, Goddard, Huang, Meng, Couch, Croll, Morris, and Ferrin]{pettersen2021ucsf}
Eric~F Pettersen, Thomas~D Goddard, Conrad~C Huang, Elaine~C Meng, Gregory~S Couch, Tristan~I Croll, John~H Morris, and Thomas~E Ferrin.
\newblock {UCSF} {C}himera{X}: {S}tructure visualization for researchers, educators, and developers.
\newblock \emph{Protein Science}, 30\penalty0 (1):\penalty0 70--82, 2021.

\end{thebibliography}

\newpage
\appendix

\section{Metric heatmaps of one-shot \textit{ab initio} runs} \label{app:sshot}
In this section, experiments with $K > M$ were also included to complete the single-shot exploration grid.

\begin{figure}[ht]
  \begin{center}
    \includegraphics[width=1.0\columnwidth, trim={0 0cm 0 0cm}, clip]{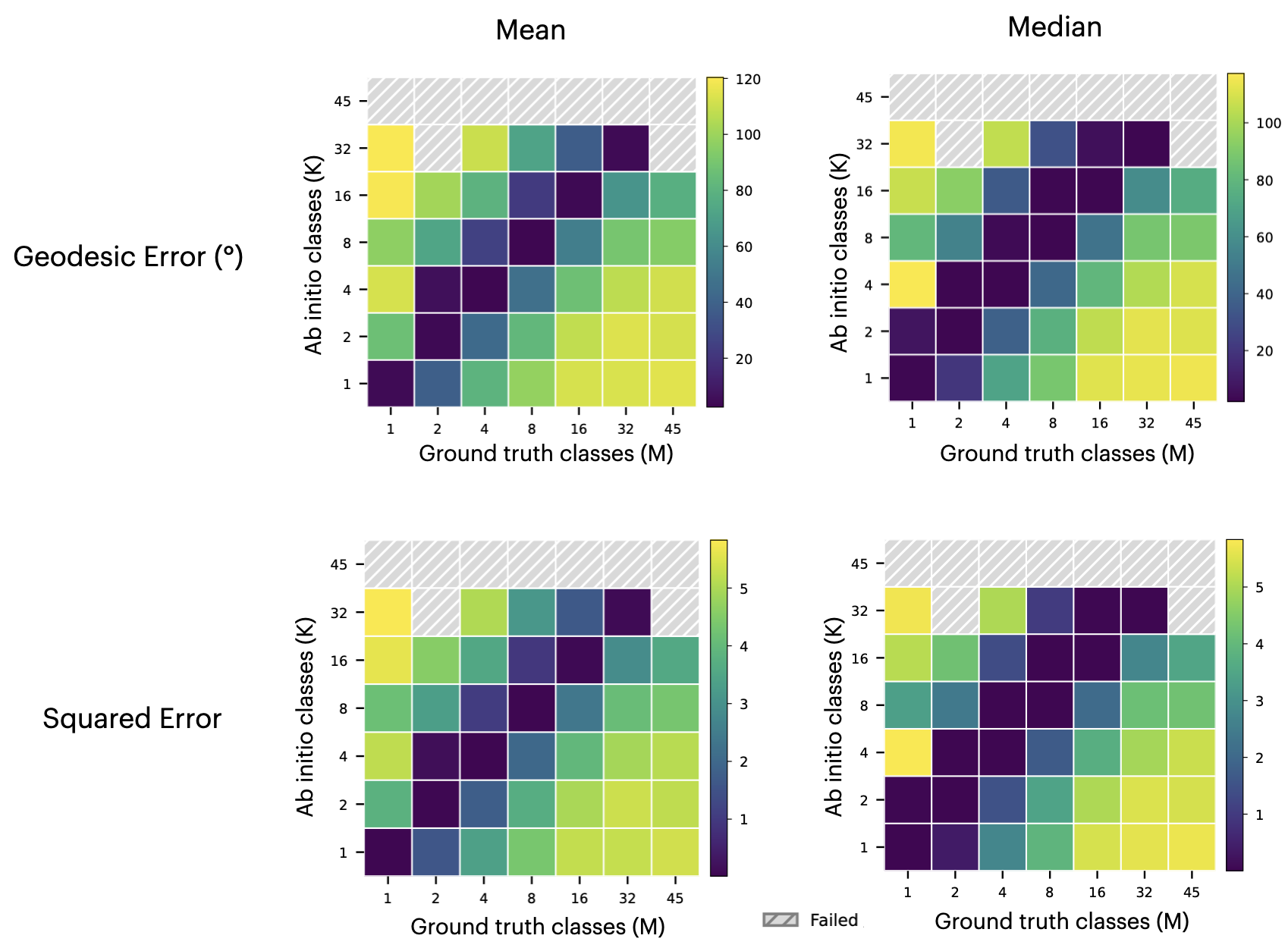}
    \caption{
      Mean (left) and median (right) pose error metrics of the single-shot \textit{ab initio} runs on subsets of Tomotwin-45-asym with increasing heterogeneity.
    }
    \label{fig:pose_err_hmap}
  \end{center}
\end{figure}

Figure~\ref{fig:pose_err_hmap} reports the mean and median pose error metrics. The geodesic error is the rotation angle (in degrees), and the squared error is the squared error between the true and predicted rotation matrices.

\begin{figure}[h]
  \begin{center}
    \includegraphics[width=1.0\columnwidth, trim={0 0cm 0 0cm}, clip]{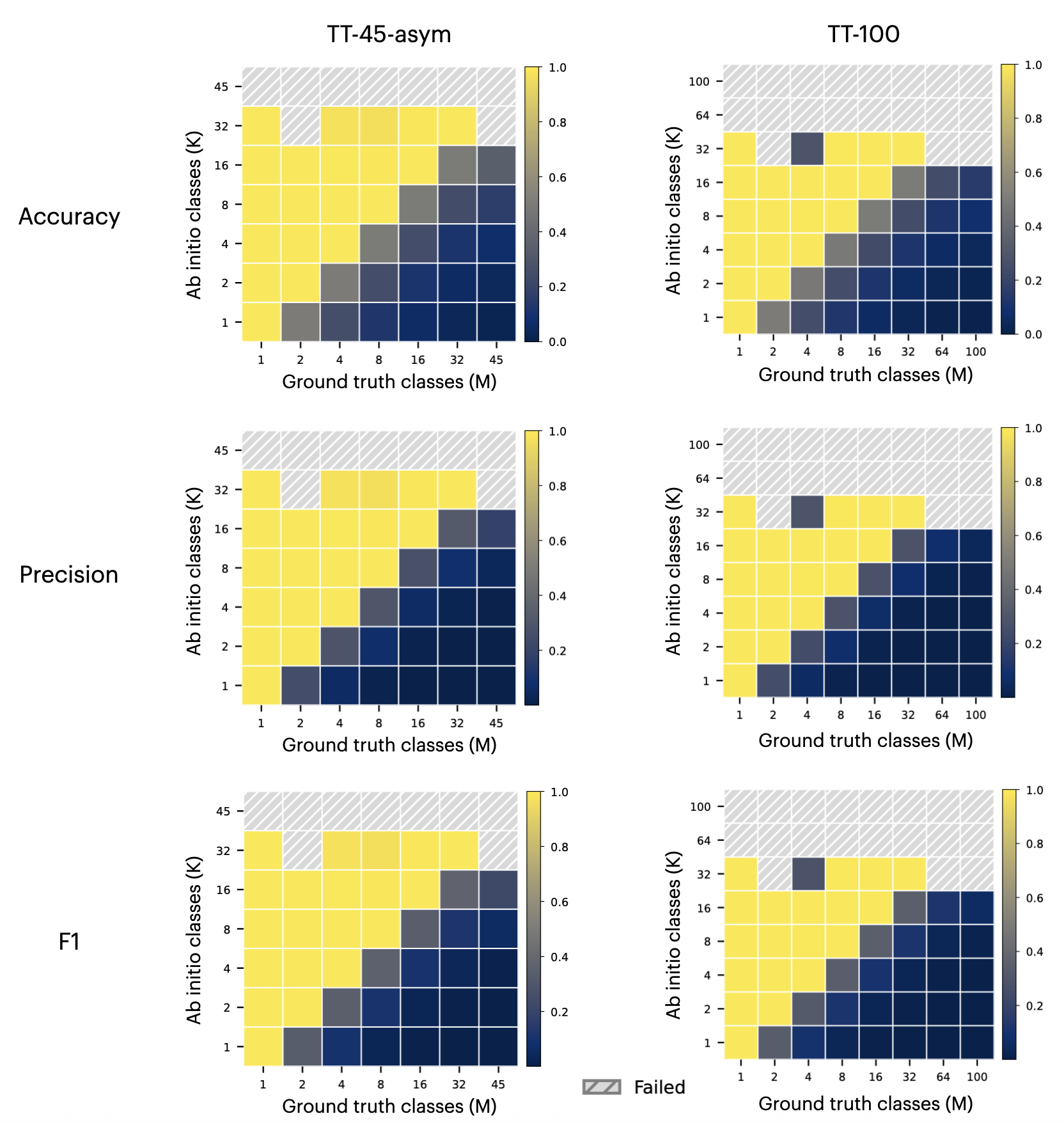}
    \caption{
      Classification metrics of the single-shot \textit{ab initio} runs on subsets of Tomotwin-45-asym (left) and Tomotwin-100 (right) with increasing heterogeneity.
    }
    \label{fig:class_err_hmap}
  \end{center}
\end{figure}

Figure~\ref{fig:class_err_hmap} reports the classification metrics (accuracy, precision, macro F1). Note that for the calculation of these metrics, we find the most abundant true label in each predicted class to map it to that true label, and in $K > M$ regime, this mapping is not one-to-one, therefore inducing a propensity for higher scores.

\begin{figure}[!htb]
  \begin{center}
    \includegraphics[height=1.5\columnwidth, trim={7cm 1cm 10cm 1cm}, clip]{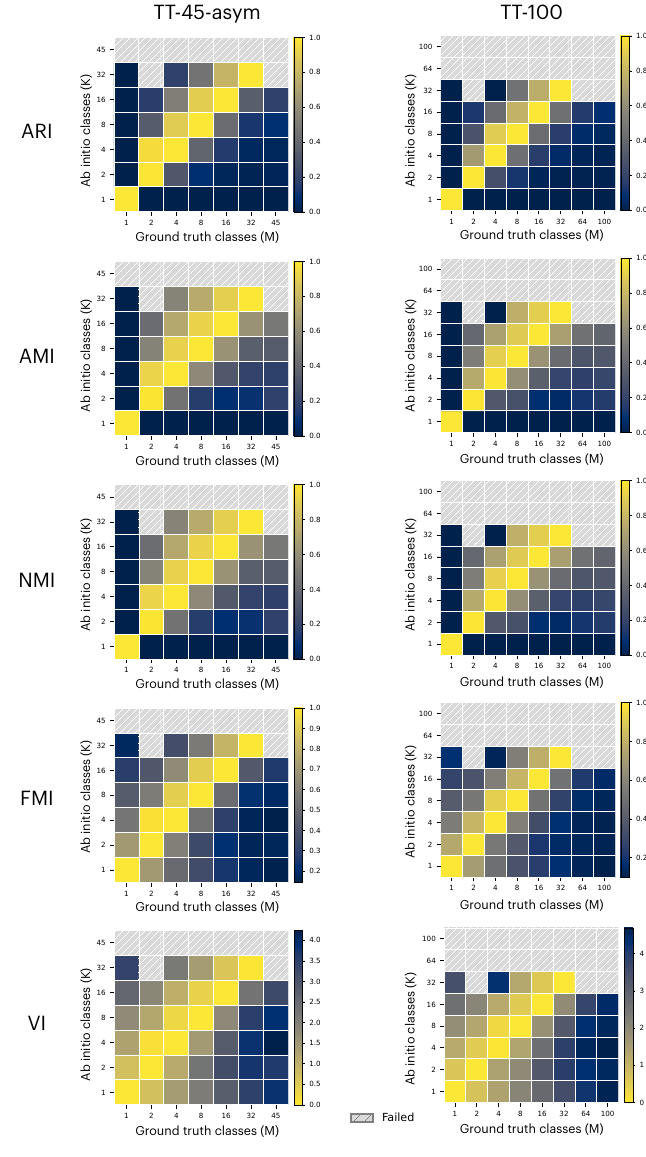}
    \caption{
      Clustering metrics of the single-shot \textit{ab initio} runs on subsets of Tomotwin-45-asym (left) and Tomotwin-100 (right) with increasing heterogeneity.
    }
    \label{fig:clust_err_hmap}
  \end{center}
\end{figure}

Figure~\ref{fig:clust_err_hmap} reports the clustering metrics calculated before the mapping mentioned above, therefore providing an alternative heuristic of the ability of the job. The metrics used are adjusted Rand index (ARI)~\cite{hubert1985ari}, adjusted mutual information (AMI)~\cite{vinh2009ami}, normalized mutual information (NMI), Fowkles-Mallows index (FMI)~\cite{fowlkes1983fmi}, and variation of information (VI)~\cite{arabie1973vi}. Out of these; ARI, AMI, and NMI are not well defined if one of the label assignments only has a single class, therefore those cells at the boundaries are set to $0$. All of these metrics were calculated using scikit-learn library~\cite{scikit-learn}.

In all figures reported in this section, failed jobs where no convergence was achieved are explicitly annotated. The results overall highlight the graceful degradation of the merit of the job as the true heterogeneity $M$ varies with respect to $K$ (i.e., off-diagonal directions).

In total, 49 jobs were executed on Tomotwin-45-asym, and 64 jobs were executed on Tomotwin-100.

\newpage

\newpage

\newpage
\section{Experimental details of the main results} \label{app:mainresults}

\subsection{Baseline methods in Tomotwin-45-asym}
Following baseline method settings were used on Tomotwin-45-asym with classification and pose accuracies as measurement of merit.

\begin{itemize}
    \item \textbf{cryoDRGN2}: The \textbf{abinit\_het} command in cryoDRGN v3.4 has been used with the canonical latent dimensionality of 8, all other parameters kept at default. The model has been trained for 20 epochs on 1 H100 in 4 hours. Using the analyze command, the resulting latent space is clustered with 45 k-means centroids, yielding the labeling subject to classification accuracy calculation.
    \item \textbf{cryoDRGN-AI}: The \textbf{train} command in DRGN-AI repository has been used with the canonical “autodecoder” architecture configuration, all parameters kept at default. The model has been trained for 111 epochs on 1 H100 in 3 hours. Using the \textbf{analyze} command, the resulting latent space is clustered with 45 k-means centroids, yielding the labeling subject to classification accuracy calculation.
    \item \textbf{Hydra}: The \textbf{train} command in Hydra repository has been used with the default configuration as DRGN-AI. Hydra was not tested on large $K$ values, and its GPU parallelization strategy does not trivially lend itself to a multi-node multi-GPU setup, therefore $K=45$ run is not viable. Facing this computational bottleneck, we trained a mixture of $K=4$, for 10 epochs, on 2 H100s, in 6 wall-hours. The resulting 4-class probability distributions of the particles were used to find the most probable class assignment at each particle.
\end{itemize}

\subsection{Hyperparameter selection}

For the results reported in Table~\ref{tab:results}, for both Tomotwin-45-asym and Tomotwin-100, the set of best performing hyperparameters in the hyperparameter ablation study (Appendix~\ref{app:hparam}) has been used:
\begin{equation}
    (\sigma, \beta, \gamma, p^{\text{min}}, n_{\text{nbr}}) = (0.6, 6, 10, 1/90, 100)
\end{equation}

\subsection{Job ensemble generation}
Recall that we define a $K$-class, $L$-level job hierarchy by a tree of $K$-class jobs, executed on their respective parent's children particle stacks in the tree structure, with a tree of height $L-1$. Therefore, one such hierarchy consists of $1 + K + \cdots + K^{L-1}$ jobs. The critical path of this tree would then include $L$ jobs (length of sequential dependency), and using the job timing assumption $t(K)=8K(\text{min})$ (Appendix~\ref{app:scaling}), the critical path would take $t_c = 8KL(\text{min})$. Assuming scalable compute access (e.g., cloud instances), the realized wall time could be as low as the critical path.

For Tomotwin-45-asym, our ensemble includes the following jobs, which are also listed on Table~\ref{tab:app_tt45asym_ensemble}:
\begin{itemize}
    \item 3 random replicates each of, 3-level hierarchies, of $K=4$.
    \item 1 random replicate of, 3-level hierarchy, of $K=8$.
    \item 4 random replicates each of, 2-level hierarchies, of $K=8$.
    \item 5 random replicates each of, 2-level hierarchies, of $K=16$.
\end{itemize}

\begin{table}[h]
    \caption{
        Job ensemble specification for the Tomotwin-45-asym analysis.
        The total number of jobs was 257, total GPU-hours used was 331.2,
        and the longest critical path was 4.3 hours.
    }
    \label{tab:app_tt45asym_ensemble}
    \centering
    \begin{tabular}{cccccc}
        \toprule
        \# Replicates & $K$ & Levels & Total Job Count & Total H100-hours & Critical path (wall-hours) \\
        \midrule
        3 & 4  & 3 & 63 & 33.6  & 1.6 \\
        1 & 8  & 3 & 73 & 77.9  & 3.2 \\
        4 & 8  & 2 & 36 & 38.4  & 2.1 \\
        5 & 16 & 2 & 85 & 181.3 & 4.3 \\
        \midrule
          &    &   & 257 & 331.2 &  4.3    \\
        \bottomrule
    \end{tabular}
\end{table}

\newpage
For Tomotwin-100, our ensemble includes the job hierarchies listed on Table~\ref{tab:app_tt100_ensemble}:

\begin{table}[h]
    \caption{
        Job ensemble specification for the Tomotwin-100 analysis.
        The total number of jobs was 215, total GPU-hours used was 308.8,
        and the longest critical path was 4.3 hours.
    }
    \label{tab:app_tt100_ensemble}
    \centering
    \begin{tabular}{cccccc}
        \toprule
        \# Replicates & $K$ & Levels & Total Job Count & Total H100-hours & Critical path (wall-hours) \\
        \midrule
        4 & 4  & 2 & 20 & 10.7  & 1.1 \\
        1 & 4  & 3 & 21 & 11.2  & 1.6 \\
        4 & 8  & 2 & 36 & 38.4  & 2.1 \\
        1 & 8  & 3 & 73 & 77.9  & 3.2 \\
        5 & 16 & 2 & 85 & 181.3 & 4.3 \\
        \midrule
          &    &   & 235 & 319.5 & 4.3 \\
        \bottomrule
    \end{tabular}
\end{table}

In our cryoSPARC setup, we had access to $\approx 50$ parallel GPU job submission lanes, therefore both of these ensembles could be done overnight.

\newpage
\section{Volume renderings of Tomotwin-45-asym results} \label{app:tt45maps}

\begin{figure}[ht]
  \begin{center}
    \includegraphics[width=0.8\columnwidth, trim={0 0cm 0 0cm}, clip]{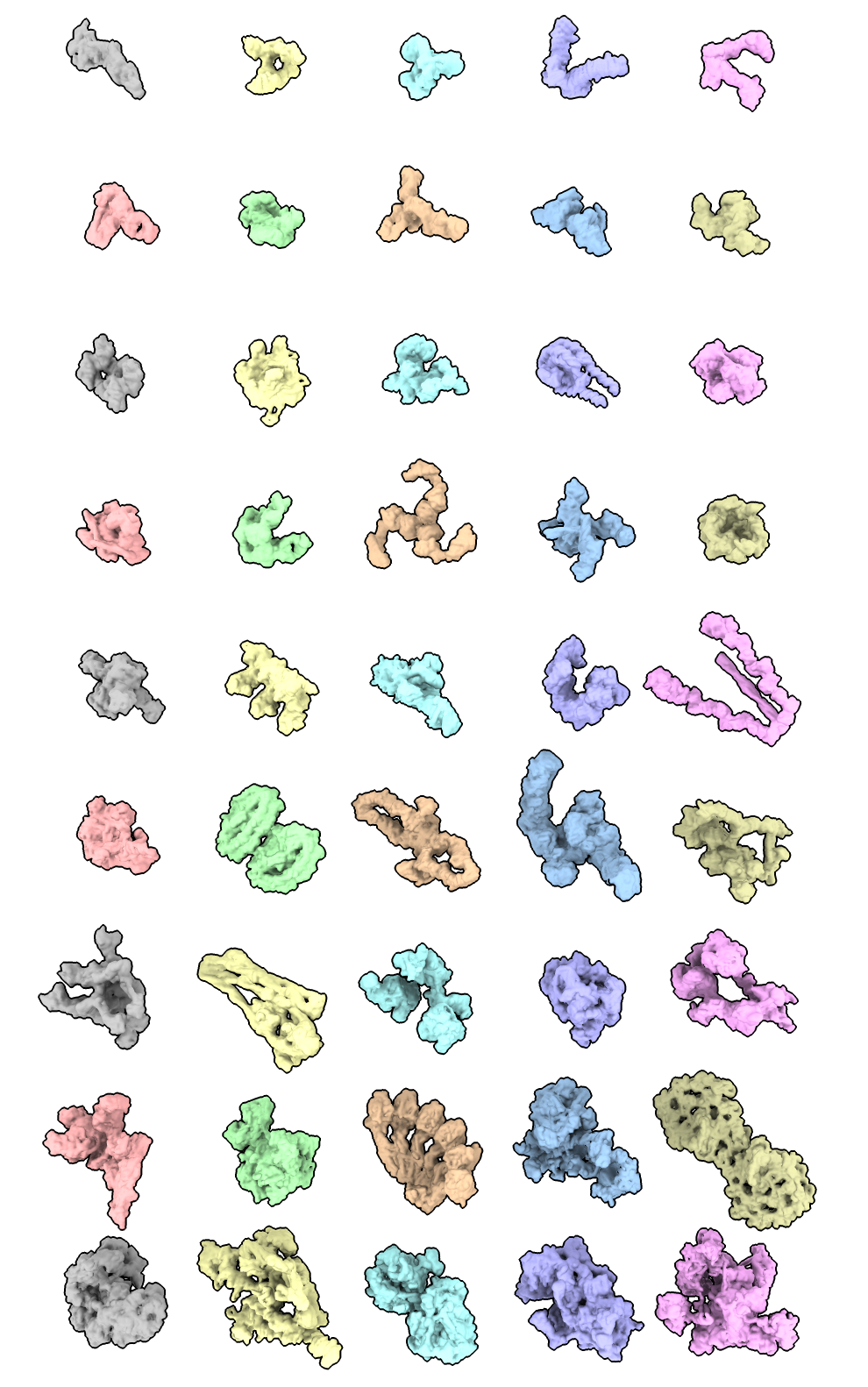}
    \caption{
      The volumes obtained from Tomotwin-45-asym ensemble classification, as reported in Table~\ref{tab:results}. The resulting particle classes were fed into homogeneous refinement to yield these refined volumes per each class, and the renderings were done in ChimeraX~\cite{pettersen2021ucsf}. 
    }
    \label{fig:tt45maps}
  \end{center}
\end{figure}

\newpage
\section{Ensemble scaling experiments} \label{app:scaling}

\begin{figure}[ht]
  \begin{center}
    \includegraphics[width=1.0\columnwidth, trim={0 0cm 6cm 4cm}, clip]{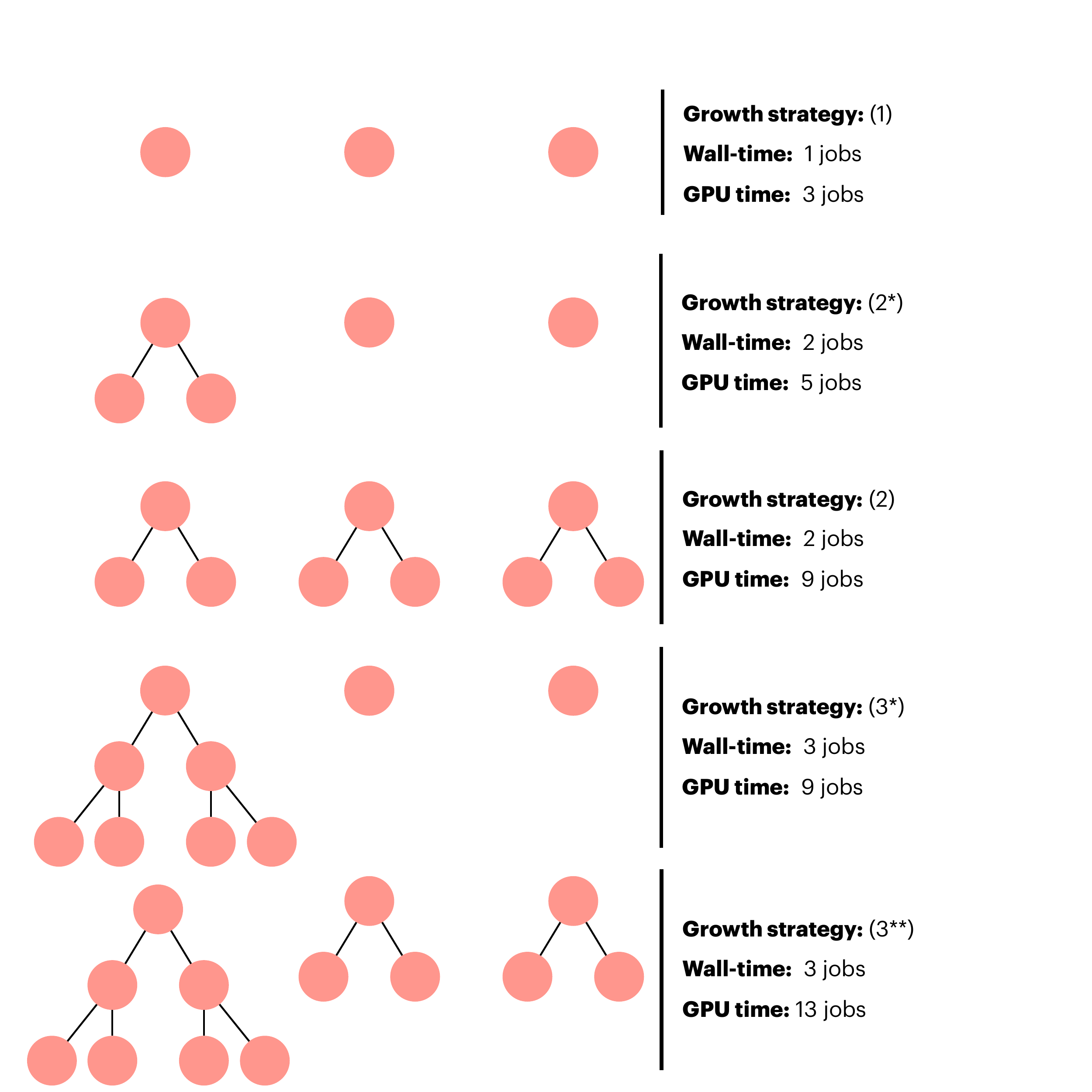}
    \caption{
      The job ensemble growth strategies employed in the scaling analysis, with the corresponding level annotations and computational costs. All strategies are demonstrated with 3 random replicates. Time is in the unit of jobs, where each job is assumed to require similar time for completion (under a constant $K$).
    }
    \label{fig:growth}
  \end{center}
\end{figure}

In generating the job forests that our method would aggregate as an ensemble, we face the task of enumerating the combinatorial search space of what a job ensemble could consist of. Even in our two-primitive formulation of the growth, and ignoring all the potential parametrizations one can utilize the modern \textit{ab initio} reconstruction tools with, we still face the questions of: (1) what depths of job trees to generate, (2) how many replicates of which depths to generate, (3) with what $K$ values to generate the constituent jobs.

As a reduction of the search space to highlight the first two questions, we therefore kept the $K$ values constant per each ensemble we analyzed in the scaling experiments. Therefore, the $K$-knob, for the purposes of the enumeration provided here, is elevated from a per-job parameter to a per-ensemble parameter. After this adjustment, the depth/width questions are highlighted as the principal axes of ensemble growth.

As a further reduction for the purposes of the scaling experiment, we only allow trees at maximum two different heights: the singular \textit{deep} tree, and the \textit{shallow} replicate trees. Therefore, in a given strategy, the \textit{level} number indicates the height+1 of the deep tree, and the number of stars indicate the height+1 of the shallow trees. This naming was chosen to highlight the fact that, assuming abundant spot compute availability for horizontal scaling, the height of the \textit{deep} tree determines the wall-time required for the completion of the job ensemble (since other trees could be executed in parallel). These ensemble growth strategies are demonstrated in Figure~\ref{fig:growth}.

The horizontal dimension of the ensemble generation is the random replicate generation. For the purposes of our enumeration, we include all the shallow trees and the deep tree in the number of random replicates $R$. For example, a 5-replicate (3**)-strategy ensemble consists of 1 deep tree of height 2, and 4 shallow trees of height 1.

In the scaling trade-off frontiers plotted in Figure~\ref{fig:scaling}, each line consists of different random replicates under a constant $K$ and strategy. The base colors (red, green, blue) describe the $K$ value of the ensembles considered, and the hue (light to dark, with increasing GPU-hours) describes the strategy employed.

For the time calculation, we observed that the execution time of the jobs scaled linearly with $K$ ($8$ minutes per $K$), hence that fixed factor is assumed: $t(K) = 8K (\text{min})$. Each job uses a single H100 GPU. The wall-times [i.e., the critical path of sequential dependency, $t_c = 8KL(\text{min})$] of the ensembles are provided in Table~\ref{tab:walltime}.

\begin{table}[ht]
  \caption{Wall-time (in hours) of different ensemble generation strategies.}
  \label{tab:walltime}
  \centering
  \begin{tabular}{r lllll}
    \toprule
    & \multicolumn{5}{c}{Growth Strategy}           \\
    \cmidrule(r){2-6}
    $K$     & 1 & 2* & 2 & 3* & 3** \\
    \midrule
    4 & 0.53 & 1.07 & 1.07 & 1.60 & 1.60 \\
    8 & 1.07 & 2.13 & 2.13 & 3.20 & 3.20 \\
    16 & 2.13 & 4.27 & 4.27 & 6.40 & 6.40 \\
    \bottomrule
  \end{tabular}
\end{table}

Scaling analyses for both Tomotwin-45-asym and Tomotwin-100 have been conducted with the ensemble aggregation hyperparameters kept constant at:
\begin{equation}
    (\sigma, \beta, \gamma, p^{\text{min}}, n_{\text{nbr}}) = (0.45, 5, 10, 1/450, 400)
\end{equation}

\newpage
\section{Reconstruction of the bacterial 50S ribosomal subunit} \label{app:50s}

\begin{figure}[h]
  \begin{center}
    \includegraphics[width=0.95\columnwidth, trim={0 4cm 0 0cm}, clip]{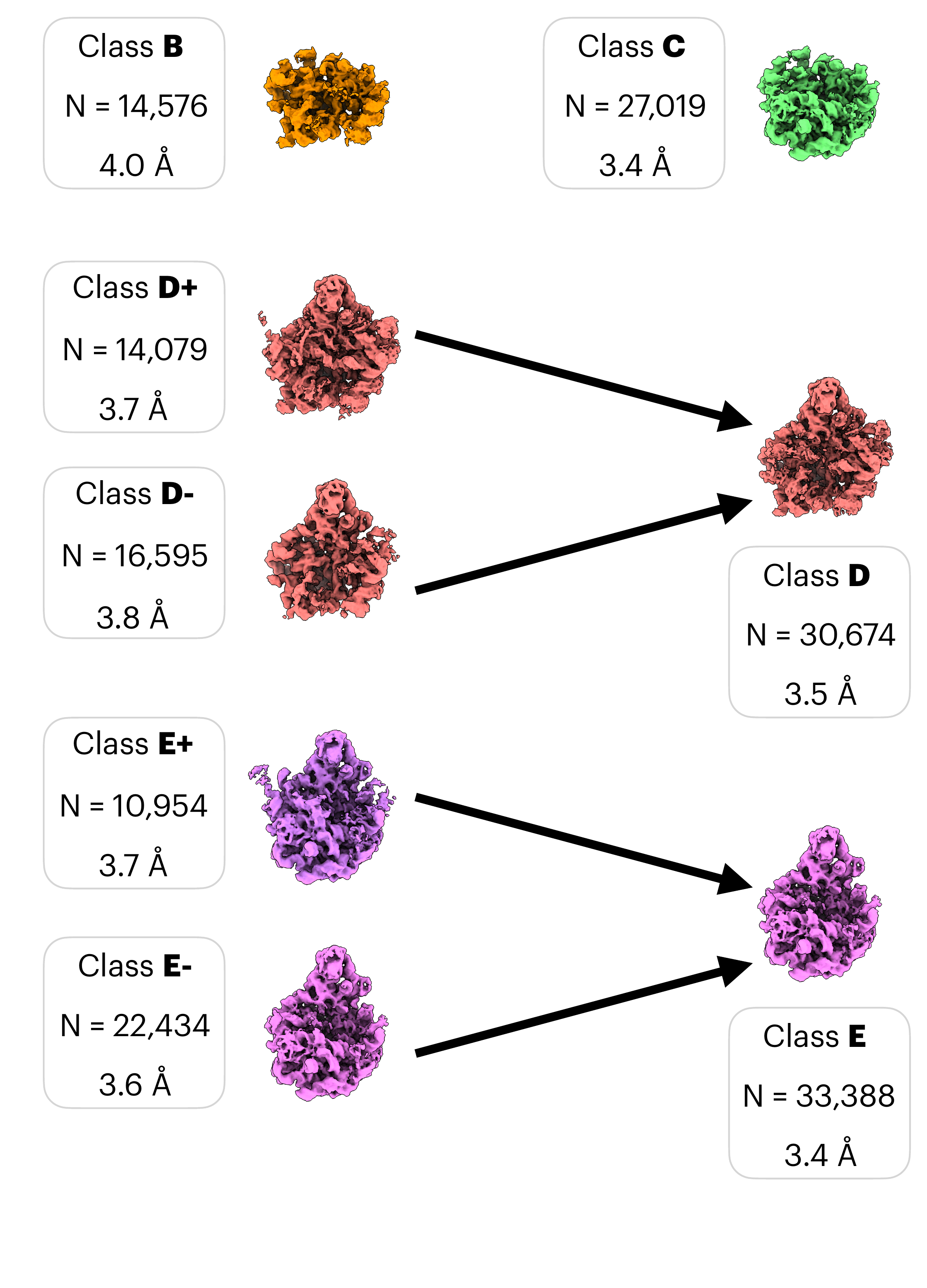}
    \caption{
    The major classes obtained through the reconstruction ensemble analysis of the 50S dataset. Class E features both the base (bottom) and the central protuberance (CP, top), Class D features only the CP, class C features only the base, and class B features none.
    }
    \label{fig:maps50s}
  \end{center}
\end{figure}

The EMPIAR-10076 entry is a single-particle cryo-EM dataset, consisting of the assembly intermediates of bL17-depleted \textit{E. coli} large ribosomal subunit (LSU)~\cite{empiar10076}. Specifically, this dataset consists of a picked particle stack ($N=132\text{k}$, $D=320\text{px}$, $1.31 \mathring{A} / \text{pix}$), including 4 major classes of LSU assembly intermediates (B, C, D, E), 2 rare classes (the small 30S subunit F, and the complete 70S ribosome A), and 1 class of junk particles. This dataset has been commonly used as a benchmark dataset for heterogeneous reconstruction and analysis methods.

\begin{wrapfigure}{R}{0.5\textwidth}
    \includegraphics[width=0.48\columnwidth, trim={0 0cm 0 0cm}, clip]{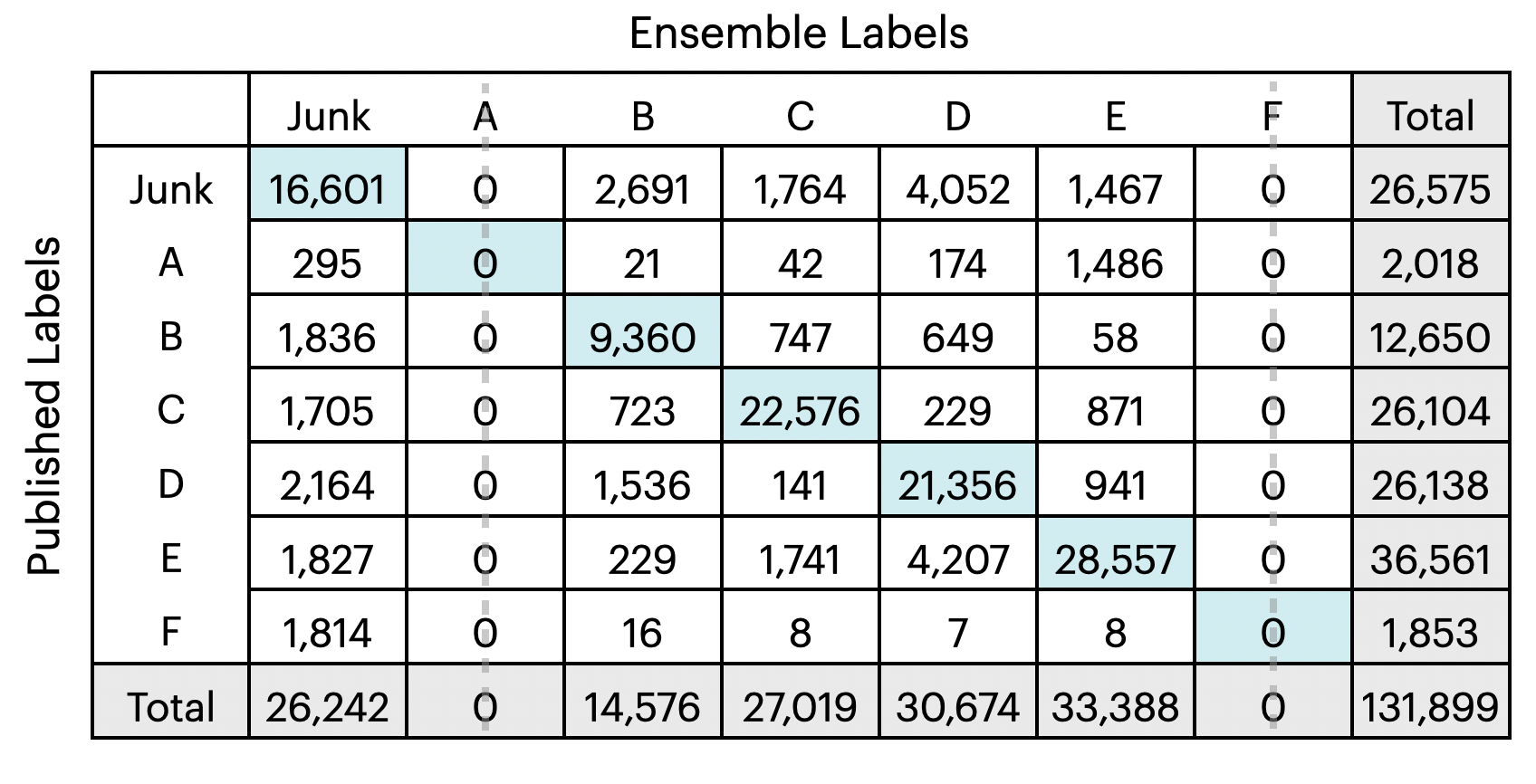}
    \caption{
    Confusion matrix between the published major class annotations and the predicted ensemble labels.
    }
    \label{fig:50s_confusion}
\end{wrapfigure}

The canonical setup of this benchmark dataset follows the cryoDRGN~\cite{zhong2021cryodrgn} filtering to reduce the stack to $N=97\text{k}$ by eliminating the junk class prior to the heterogeneity analysis. The merit of the benchmarked method, then, is determined by its success in isolating the 4 major classes.

In order to provide stronger evidence of practical applicability of our method, we expand on this setup by processing the whole unfiltered stack $(N=132\text{k})$, directly downloadable from EMPIAR. The major state clasification is done \textit{ab initio}, i.e.~without prior pose information, in our setup. This is a more challenging setup than the known-pose/consensus volume setup used for models such as~\cite{zhong2021cryodrgn} and~\cite{chen2021deep}.

In our major state classification (Figures~\ref{fig:bar50s},~\ref{fig:maps50s}, and~\ref{fig:50s_confusion}), we reach 75\% accuracy in matching the published major class annotations (Fig. 4 of~\cite{empiar10076}), successfully isolating the dataset into 5 classes (4 major classes and the junk particle class). 

\begin{figure}[H]
  \begin{center}
    \includegraphics[page=1, width=0.95\columnwidth, trim={1.5cm 4cm 5cm 1cm}, clip]{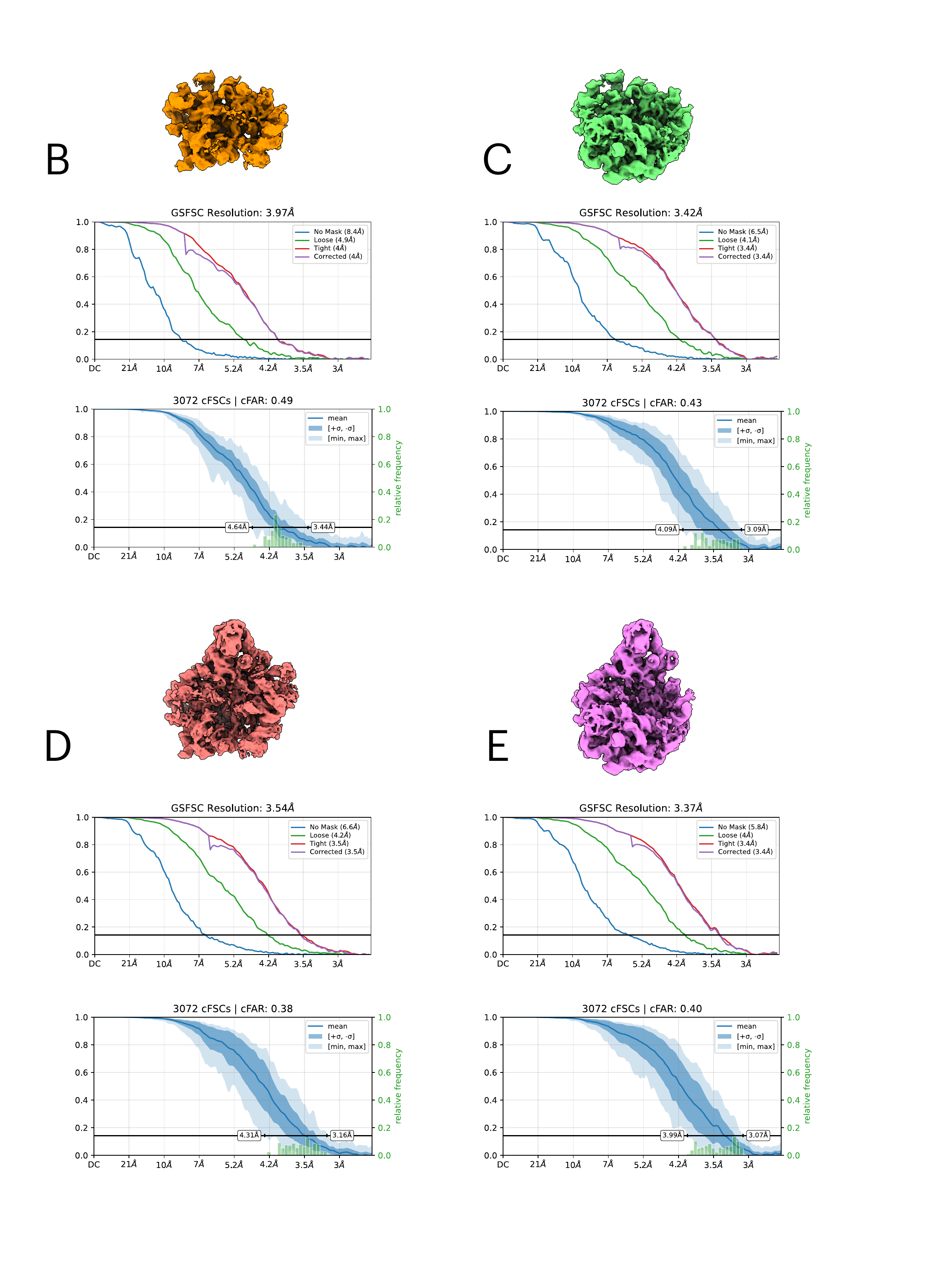}
    \caption{
    Fourier Shell Correlation (FSC) and conical FSC (cFSC) curves reported by cryoSPARC for classes B, C, D, and E.
    }
    \label{fig:fsc50s_1}
  \end{center}
\end{figure}

\begin{figure}[H]
  \begin{center}
    \includegraphics[page=2, width=0.95\columnwidth, trim={1.5cm 4cm 5cm 1cm}, clip]{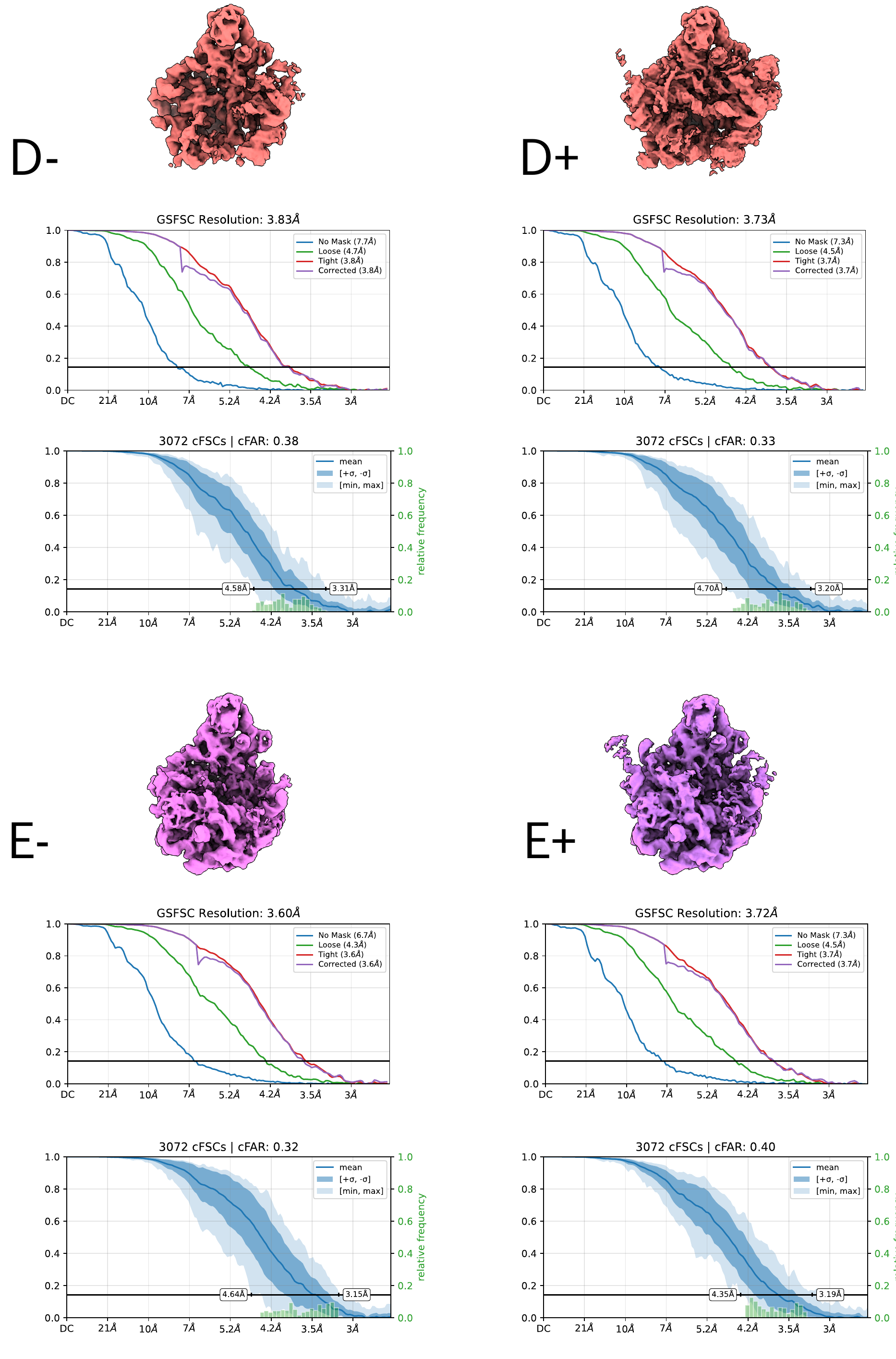}
    \caption{
    Fourier Shell Correlation (FSC) and conical FSC (cFSC) curves reported by cryoSPARC for classes D-, D+, E-, and E+.
    }
    \label{fig:fsc50s_2}
  \end{center}
\end{figure}

At the clustering stage, we used $7$ clusters with the hopes that the rare classes A and F could also be captured. Our method could not capture these rare states. However, the clusters found included a subclassification of the classes $D$ and $E$, which are named with $+$ and $-$ (Figure~\ref{fig:maps50s}). The $+$ variants feature density in the L1 stalk (left in the provided projections) and shoulder (right in the provided projections) domains, while the $-$ variants do not feature L1 stalk. This finding was consistent with the minor classifications reported in~\cite{empiar10076}. These variants were merged to produce the $5$-class classification, including the junk class (not pictured). The half-map FSC resolution and other map quality metric curves of each obtained class are reported through Figures~\ref{fig:fsc50s_1}-\ref{fig:fsc50s_2}.

In obtaining the result, our ensemble includes the following jobs, which are also listed on Table~\ref{tab:app_50s_ensemble}:
\begin{itemize}
    \item 3 random replicates each of, 2-level hierarchies (root and children jobs, i.e.~$K+1$ jobs total in a tree), of $K=\{4, 8, 16\}$.
    \item 3 random replicates each of, 3-level hierarchy (root, children, and grandchildren jobs, i.e.~$K^2+K+1$ jobs total in a tree), of $K=6$.
\end{itemize}

\begin{table}[h]
    \caption{
        Job ensemble specification for the 50S analysis.
        The total number of jobs was 222, total GPU-hours used was 248.8,
        and the longest critical path was 4.3 hours.
    }
    \label{tab:app_50s_ensemble}
    \centering
    \begin{tabular}{cccccc}
        \toprule
        \# Replicates & $K$ & Levels & Total Job Count & Total H100-hours & Critical path (wall-hours) \\
        \midrule
        3 & 4  & 2 & 15  & 8.0   & 1.1 \\
        3 & 8  & 2 & 27  & 28.8  & 2.1 \\
        3 & 16 & 2 & 51  & 108.8 & 4.3 \\
        3 & 6  & 3 & 129 & 103.2 & 2.4 \\
        \midrule
          &    &   & 222 & 248.8 & 4.3 \\
        \bottomrule
    \end{tabular}
\end{table}

Assuming scalable compute access (e.g., cloud instances), the wall time could be as convenient as 4.5 hours. In our cryoSPARC setup, we had access to $\approx 50$ parallel GPU job submission lanes, therefore our experiments were done overnight. Finally, the aggregation hyperparameters used were:
\begin{equation}
    (\sigma, \beta, \gamma, p^{\text{min}}, n_{\text{nbr}}) = (0.45, 5, 10, 1/450, 400)
\end{equation}

\newpage

\newpage
\section{Hyperparameter tuning study} \label{app:hparam}
We pursue a two-stage approach in our 5-dimensional hyperparameter grid sweep for alleviating the curse of dimensionality. We split the hyperparameters into two sets: the exponential factors $(\sigma, \beta, \gamma)$ in the weight heuristics, and the other two $(p^{\text{min}}, n_{\text{nbr}})$, namely the tiny class fraction threshold and the number of neighbors to keep in the sparsification process. In the following discussion, whenever $\%$ is used with respect to accuracy variation, it represents absolute percentage points and not relative variations.

\begin{wrapfigure}{R}{0.5\textwidth}
    \includegraphics[width=0.48\columnwidth, trim={0 0cm 0 0.8cm}, clip]{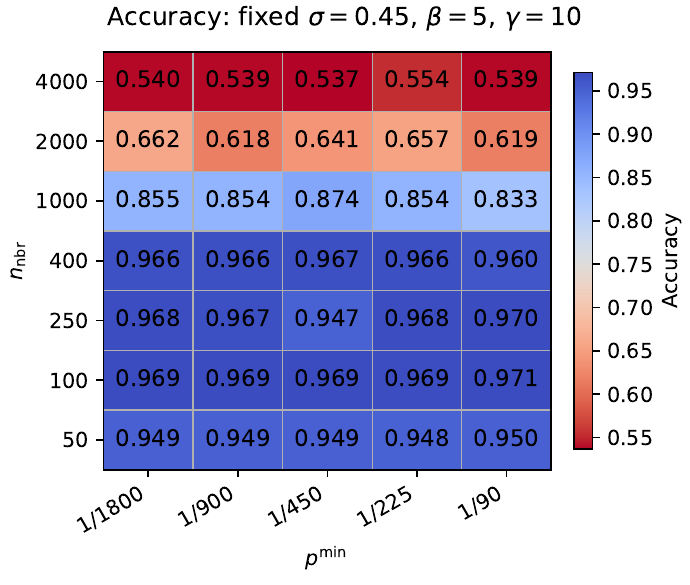}
    \caption{
    Accuracies obtained in the hyperparameter grid sweep for the tiny class proportion threshold $p^{\text{min}}$ and the number of nearest neighbors kept $n_{\text{nbr}}$ in the sparsification.
    }
    \label{fig:hparam_tn}
\end{wrapfigure}

We start with $(\sigma, \beta, \gamma) = (0.45, 5, 10)$ and conduct a grid sweep on $(p^{\text{min}}, n_{\text{nbr}})$, reported in Figure~\ref{fig:hparam_tn}. The results show negligible dependency on the tiny class proportion ($< 4 \%$ fluctuation), but a crucial dependency on the sparsification threshold ($\approx 45 \%$ change). Specifically, we observe that sparsifying the dense matrix up to only a few hundreds of neighbors remaining is important. Note that this is encouraging from a computational complexity perspective: the more the similarity matrix is sparsified, the less time and memory complexity the clustering methods would require (reduction of complexity from quadratic to linear in terms of the number of particles).

\begin{figure}[ht]
  \begin{center}
    \includegraphics[width=1\columnwidth, trim={0 0cm 0 1cm}, clip]{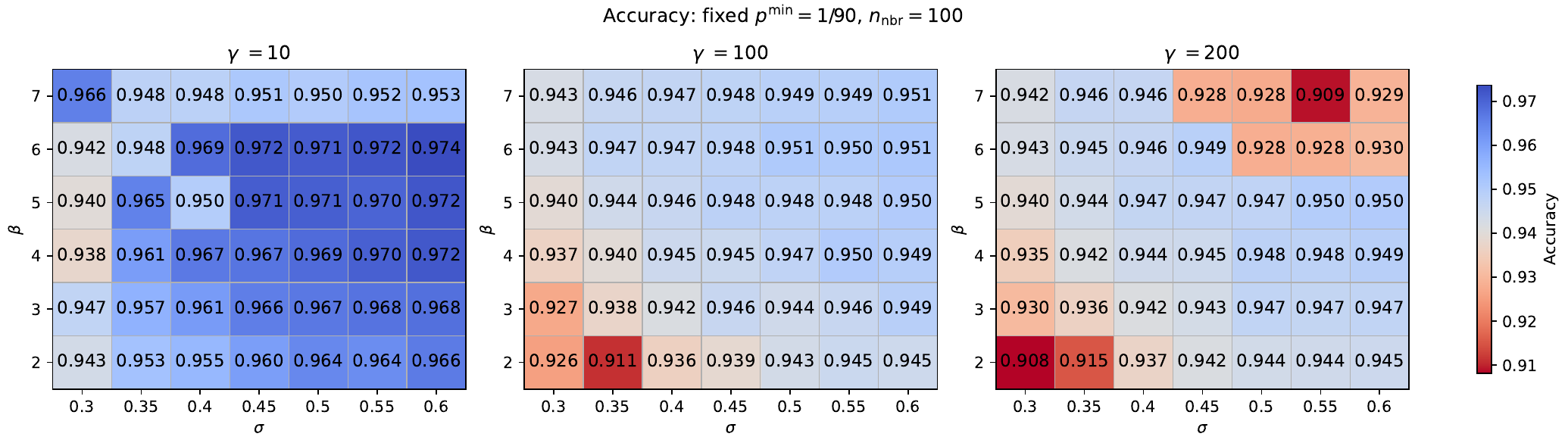}
    \caption{
    Accuracies obtained in the hyperparameter grid sweep for the exponential rate factors $(\sigma, \beta, \gamma)$ in the weight heuristics.
    }
    \label{fig:hparam_sgb}
  \end{center}
\end{figure}

In the second phase of the hyperparameter search, we start with the result of the first sweep and explore $(\sigma, \beta, \gamma)$. Specifically, we pick the best cell from above, i.e. $(p^{\text{min}}, n_{\text{nbr}}) = (1/90, 100)$. The results are demonstrated in Figure~\ref{fig:hparam_sgb}. We observe that the fluctuations are limited to a band of $10 \%$, highlighting the robustness of the weight heuristics under no tuning, while showing upside potential in tuning for minor performance gains in more detailed analyses.

For the results reported in Table~\ref{tab:results}, for both Tomotwin-45-asym and Tomotwin-100, the set of best performing hyperparameters above has been used:
\begin{equation}
    (\sigma, \beta, \gamma, p^{\text{min}}, n_{\text{nbr}}) = (0.6, 6, 10, 1/90, 100)
\end{equation}

The job ensembles used in this study are the same as in the main results, therefore requiring no further GPU-hours. The primary driver of the computational complexity in this hyperparameter sweep is the clustering step. Each clustering assignment takes a few minutes on a single CPU core, hence using parallel CPUs, the ablation study is done in an hour.


\newpage

\end{document}